\documentclass[11pt, letterpaper]{article}
\usepackage{epsfig,mst-stylefile,amssymb,amsmath, multirow, url, tcolorbox,booktabs, enumitem, xcolor}
\usepackage{pdflscape}
\usepackage[ruled]{algorithm2e}
\usepackage{kotex}
\usepackage{cleveref}

\usepackage{natbib}
\renewcommand{\baselinestretch}{1.15}
\graphicspath{{./}}

\newtheorem{@assumption}{\sc Assumption}[section]

\begin{document}

\title{ A Korean Macroeconomic Database for \\ Data-Rich Policy Analysis and U.S.--Korea Dependence
\footnote{All authors contributed equally in this work.}
\footnote{Changryong Baek was supported by the National Research Foundation of Korea grant funded by the Korea government (MSIT) (RS-2025-00519717).}
\footnote{Seung Hyun Moon's research was supported by Basic Science Research Program through the National Research Foundation of Korea (NRF) funded by the Ministry of Education(RS-2025-25414103).}}

\author{
Changryong Baek\footnote{Department of Statistics, Sungkyunkwan University, 25-2, Sungkyunkwan-ro, Jongno-gu, Seoul, Korea 03063, crbaek@skku.edu}\\Sungkyunkwan University \and Seung Hyun Moon\footnote{Department of Statistics, Seoul National University, 1 Gwanak-ro, Gwanak-gu, Seoul, Korea, msh94kr@snu.ac.kr}\\Seoul National University
\and Seunghyeon Lee\footnote{Bank of Korea, 39 Namdaemun-ro, Jung-gu, Seoul, Korea 04531, seunghyeon.lee@bok.or.kr} \\ Bank of Korea
}

\date{\today}

\maketitle

\begin{abstract}
We introduce KRED (Korea Research Economic Database), a FRED-MD-compatible monthly macroeconomic database for Korea designed for data-rich policy analysis and cross-country comparison. KRED contains 125 monthly series from ECOS, KOSIS, and administrative labor-market sources, with coverage back to 1960. Using monthly data over 2009:06--2025:12, we construct four common factors that summarize interest rates and term spreads, industrial activity, housing permits, and the labor market. Their factor-based diffusion indices trace the different timing of interest-rate, industrial, housing, and labor-market developments in Korea.
We then use KRED in two empirical applications. First, a factor-augmented VAR shows that U.S. monetary tightening weakens Korean real activity and is followed by an increase in the Korean policy rate. The broader KRED information set also substantially attenuates the anomalous positive CPI price-level response generated by a small VAR, and the reductions in the average and peak responses remain robust across bootstrap designs that allow for residual dependence. Second, a grouped U.S.--Korea tensor autoregression shows that cross-country dependence is concentrated in financially oriented blocks. Transmission from U.S. financial conditions to Korea is stronger than in the reverse direction, whereas spillovers through real activity and housing are weaker and less persistent.
KRED thus provides a transparent public database for Korean macroeconomic research and a useful foundation for policy analysis and comparative work on macro-financial dependence in Asia.
\end{abstract}

\clearpage

\section{Introduction} \label{s:intro}

Large macroeconomic panels have become standard inputs in empirical macroeconomics. Early data-rich studies showed that large sets of indicators improve forecasting and help summarize aggregate fluctuations, but this line of work often relied on datasets that required substantial manual curation. FRED-MD, developed by \citet{mccracken2015fredmd}, changed this practice by providing a public monthly database with standardized transformations and documentation, and it has become a leading benchmark for transparent and reproducible macroeconomic research. Related efforts have also emerged outside the United States. For Canada, \citet{fortingagnon2022large} provide a large macroeconomic database with public real-time vintages; for the euro area and its member countries, \citet{barigozzi2024large} develop a harmonized dataset for comparative macroeconomic and policy analysis; and for Korea, \citet{kimandswanson2018} assemble real-time macroeconomic data for GDP backcasting, nowcasting, and forecasting. These efforts differ in scope: the Canadian and euro-area databases provide broad research infrastructures, whereas the existing Korean contribution is organized primarily around real-time GDP backcasting, nowcasting, and forecasting.

This paper constructs KRED, a FRED-MD-compatible monthly macroeconomic database for Korea. Its primary purpose is to provide a transparent and reusable foundation for data-rich policy analysis in a small open economy that is highly exposed to external monetary and financial conditions. This perspective is closely related to the broader small-open-economy literature that emphasizes the importance of foreign disturbances and external financial conditions for domestic dynamics; see, for example, \citet{justiniano2010smallopen,rey2015dilemma}. It is also directly connected to the literature on the international transmission of U.S. monetary-policy shocks, beginning with \citet{kim2001usmp}. For Korea and Asia, related evidence shows that external monetary conditions matter for Korean capital flows and trade and for the lending behavior of Asian banks more generally; see \citet{ree2014safehaven} and \citet{lee2022spilloversasia}. At the same time, close alignment with the FRED-MD architecture makes KRED useful for comparative work and for cross-country models that combine Korea with other economies in a standardized way. Such standardization is particularly valuable when country-specific panels are brought into multicountry VARs and more structured matrix- and tensor-based dynamic models; see, for example, \citet{canova2009estimating,canova2013panel,hill2021tensor,li2021multilinear,tsay2024matrix,wang2024tensorar,luo2025bayesian}. In this sense, KRED is designed not only as a domestic macroeconomic database for Korea but also as an empirical infrastructure for studying international macro-financial dependence.

Constructing such a database for Korea is nontrivial because relevant series are dispersed across multiple public platforms rather than provided through a single integrated system. KRED consolidates data from the ECOS system of the Bank of Korea, the KOSIS portal from the Ministry of Data and Statistics, and administrative labor-market statistics from the Ministry of Employment and Labor within a reproducible workflow. To facilitate transparency, replication, and future empirical work, the data repository is publicly available at \url{https://github.com/crbaek/KRED} and will be updated regularly. The initial release contains 125 monthly series dating back to 1960:01. Over the common empirical sample, a four-factor representation based on the leading principal-component space summarizes interest rates and term spreads, industrial activity, housing permits, and the labor market. The corresponding diffusion indices reveal that these dimensions of the Korean economy evolve at different speeds and provide a compact account of major macroeconomic episodes.

The empirical analysis proceeds in three steps. We first document the four-factor structure and use the factor-based diffusion indices to trace the sequencing of interest-rate, industrial, housing, and labor-market developments. We then estimate a factor-augmented vector autoregression (FAVAR)
following \citet{bernanke2005measuring}, which shows that U.S. monetary
tightening transmits to Korean activity and policy rates. Relative to a small VAR, the broader KRED information set substantially attenuates the anomalous positive CPI price-level response, with robust reductions in its average and peak magnitudes. Finally, because KRED shares a common grouped architecture with FRED-MD, we estimate a grouped U.S.--Korea tensor autoregression and find that cross-country dependence is concentrated in financially oriented blocks, with stronger transmission from the U.S. financial block to Korea than in the reverse direction.

These results position KRED as more than a data archive. It provides a transparent public database for Korean macroeconomic research, a practical input for data-rich policy analysis, and a standardized building block for comparative work on cross-country macro-financial dependence. This combination is particularly relevant in an Asian setting, where external monetary conditions and cross-border financial linkages are central to the interpretation of domestic macroeconomic fluctuations.

The remainder of the paper is organized as follows. Section~\ref{s:data} describes the construction of KRED and the main design choices. Section~\ref{s:empirical-analysis} presents the factor analysis and factor-based diffusion indices. Section~\ref{s:favar} studies monetary-policy shocks and the price-puzzle comparison. Section~\ref{s:tenar} examines grouped U.S.--Korea dependence. Section~\ref{s:conclusion} concludes. The Appendix reports implementation details, a supplementary tensor impulse-response contrast, FAVAR robustness checks, and the KRED/FRED-MD variable mapping.

\section{KRED construction} \label{s:data}

KRED serves two related purposes: it is a general-purpose monthly macroeconomic database for Korea, and it is a standardized country block aligned as closely as possible with FRED-MD. This section documents the data sources, grouping and transformation rules, and the main departures required by Korean institutional features.

Our data come from three public sources: the Economic Statistics System (ECOS) of the Bank of Korea (\url{https://ecos.bok.or.kr/}), the KOSIS portal (\url{https://kosis.kr/index/index.do}) from the Ministry of Data and Statistics, and employment statistics from the Ministry of Employment and Labor (\url{https://laborstat.moel.go.kr/}). KRED contains 125 time series. Variables are organized into eight groups: (i) output and income, (ii) labor market, (iii) housing, (iv) consumption, orders, and inventories, (v) money and credit, (vi) interest rates and exchange rates, (vii) prices, and (viii) the stock market. Transformation codes follow FRED-MD as closely as possible, and Appendix~\ref{appendix:KREDvariables} reports both the transformation codes and the series-level mapping to FRED-MD counterparts. The static-factor and FAVAR applications use distinct but explicitly related information sets, described below.

While the overall structure aligns with FRED-MD, several important differences reflect the unique characteristics of Korean macroeconomic data:

\begin{itemize}
	
	\item \textbf{Labor Market}: The FRED-MD Help Wanted Index (\texttt{HWI}) is replaced with the monthly number of newly registered job openings in KRED. Similarly, \texttt{HWIURATIO} is substituted with the job openings-to-seekers ratio, which reflects the average number of available jobs per job seeker. FRED-MD includes high-frequency indicators like weekly unemployment insurance claims, whereas KRED relies on monthly data. For instance, the U.S.\ category ``Unemployed for 5--14 weeks'' is approximated by Korea's ``Unemployed less than 3 months'' series. Because the Korean duration statistics combine interval and cumulative categories, we form mutually exclusive bins before computing the mean; in particular, $U_{6\text{--}12}=U_{\ge6}-U_{\ge12}$. We then approximate the mean unemployment duration (\texttt{UEMPMEAN}) in months by
	\begin{equation} \label{e:uempmean-formula}
		\mathrm{\texttt{UEMPMEAN}}
		= \frac{1.5\,U_{<3} + 4.5\,U_{3\text{–}6} + 9\,U_{6\text{–}12} + 12\,U_{\ge 12}}{U_{\mathrm{tot}}},
	\end{equation}
	where $U_{<3}$, $U_{3\text{–}6}$, $U_{6\text{–}12}$, and $U_{\ge 12}$ denote the numbers of unemployed people with durations less than 3 months, 3 to 6 months, 6 to 12 months, and at least 12 months, respectively, and $U_{\mathrm{tot}}=U_{<3}+U_{3\text{–}6}+U_{6\text{–}12}+U_{\ge 12}$.
	
	\item \textbf{Housing Market}: FRED-MD provides regional disaggregation of housing starts and permits by U.S.\ census regions. KRED refines this by categorizing housing data based on Korea’s urban structure: Seoul, the Seoul metropolitan area (Incheon and Gyeonggi), five major cities (Busan, Daegu, Daejeon, Gwangju, Ulsan), and other regions. This enhances the regional granularity of housing dynamics.
	
	\item \textbf{Interest Rates and Yields}: Short-term Korea Treasury Bills (KTBs) markets remain less developed than those in the United States. For short horizons, we therefore use Monetary Stabilization Bond yields at 91 days and 1 year as practical counterparts to the U.S. T-bills.\footnote{Monetary Stabilization Bonds---issued by the Bank of Korea to manage liquidity surplus---serve as reference rates for short maturities in Korea.} For the term structure, we construct spreads relative to the policy rate. In our empirical analysis, the 3-year spread is particularly informative for Korea, and we use the 3-year term spread (\texttt{T3YFFM}) as the primary medium-maturity slope measure.
	
	\item \textbf{Exchange Rates}: KRED includes exchange rates of the Korean Won against major currencies, including the U.S.\ dollar, euro, Japanese yen, and Chinese yuan. These selections are based on Korea's export shares by trade partner, providing relevant indicators of Korea’s external balance and competitiveness.
	
	\item \textbf{Real Personal Income}: Korea does not publish a BEA-style real personal income (\texttt{RPI}) series. Although household disposable income is available, a directly comparable pre-tax personal income measure for the household sector (households+NPISH) is not. In the United States, the \texttt{RPI} series we reference is already expressed in per-capita terms, so multiplying by population yields an aggregate measure of total real personal income. Since an analogous personal-income aggregate is unavailable for Korea, we proxy this concept using economy-wide aggregates, namely seasonally adjusted quarterly real \texttt{GDP} and real \texttt{GNI}, which provide a consistent summary of Korea’s macroeconomic activity. These proxies differ from \texttt{RPI} in both concept (economy-wide rather than household-sector income) and frequency (quarterly rather than monthly), but they are standard, consistently measured, and adequate for our purpose. To construct quarterly real \texttt{GDP} and real \texttt{GNI} series from annual data and quarterly growth rates, we apply the Denton method \citep{denton1971adjust}. The resulting quarterly paths are subsequently linearly interpolated to monthly frequency on the level scale before the transformation codes are applied, following the common preprocessing rule described in Section~\ref{s:data-preprocessing}. Further details on the quarterly benchmarking step are provided in Appendix~\ref{apdx:denton-method}.
	
	\item \textbf{Production Index}: Since Korea's industrial production does not include aggregates directly comparable to \texttt{IPFPNSS} and \texttt{IPFINAL} in the BEA, we construct proxy indices to approximate them. For \texttt{IPFPNSS}, we combine construction (\texttt{IPFPNSS1}) and service-industry (\texttt{IPFPNSS2}) production indices from KOSIS, reflecting the sectoral scope embedded in the corresponding FRED series. For \texttt{IPFINAL}, we use the capital-goods (\texttt{IPFINAL1}) and consumer-goods (\texttt{IPFINAL2}) indices from KOSIS to mirror the composition of the original measure.
	
	\item \textbf{Series excluded relative to FRED-MD}: A small number of FRED-MD series have no feasible monthly analogue in Korea: \texttt{CMRMTSPx} is available only at annual frequency in ECOS; Total business inventories (\texttt{BUSINVx}) are unavailable even though the inventories-to-sales ratio (\texttt{ISRATIOx}) can be constructed; \texttt{CUSR0000SA0L5} is omitted because \texttt{CPIAUCSL} and \texttt{CPIMEDSL} are available only as index series and the expenditure weights needed to remove medical care are not available.
	
	\item \textbf{Seasonal adjustment}: We use seasonally adjusted series when an official adjustment is available. When it is not, the database preserves the source series rather than applying an undocumented adjustment. All transformations and empirical overrides used below are recorded in the replication files.
\end{itemize}

\subsection{Empirical samples and preprocessing}\label{s:data-preprocessing}

The empirical sample runs from 2009:06 to 2025:12, giving $T=199$ monthly observations. The static-factor analysis uses the 103 KRED series with complete coverage over this window. The FAVAR information set adds \texttt{US.FFR} as an observed external-policy variable and therefore contains 104 series; \texttt{US.FFR} is not included in the Korean static-factor extraction.

The ten quarterly series recorded as repeated quarterly values, including the Denton-constructed real GDP and real GNI series, are linearly interpolated to monthly frequency on the level scale before their transformation codes are applied; the final segment is extrapolated using the preceding quarterly slope. No interpolation is performed after transformation, and the retained transformed static panel contains no missing observations. The three PCE component series \texttt{DDURRG3M086SBEA}, \texttt{DNDGRG3M086SBEA}, and \texttt{DSERRG3M086SBEA} are already annualized growth rates and are therefore used with transformation code 1 rather than differenced again.

The FAVAR application uses a small number of documented analytical overrides. \texttt{KR.MIR}, \texttt{UNRATE}, and \texttt{US.FFR} enter in levels, while \texttt{CPIAUCSL} enters as a log difference. These overrides apply to the FAVAR information set and do not alter the general-purpose transformation codes reported in Appendix~\ref{appendix:KREDvariables}.

\section{KRED empirical analysis} \label{s:empirical-analysis}

\subsection{Factor extraction and the number of factors}\label{s:factor-number}

To summarize the common variation in KRED, we use a static approximate factor model for the balanced transformed Korean panel described in Section~\ref{s:data-preprocessing}. Let $q$ denote its cross-sectional dimension, let $x_t\in\mathbb{R}^{q}$ be the vector observed at month $t$, and stack the panel as $X=(x_1,\ldots,x_T)\in\mathbb{R}^{q\times T}$. The model is
\[
X=\Lambda F+E,
\qquad
\Lambda\in\mathbb{R}^{q\times r},\quad F\in\mathbb{R}^{r\times T}.
\]
After centering and standardizing each series, let $S=T^{-1}XX'$ and write its ordered eigenvalues as $d_1\ge\cdots\ge d_q\ge0$. With $V_r$ collecting the first $r$ eigenvectors, the principal-components estimators are
\[
\widehat\Lambda(r)=\sqrt q\,V_r,
\qquad
\widehat F(r)=q^{-1}\widehat\Lambda(r)'X,
\]
which satisfy $q^{-1}\widehat\Lambda(r)'\widehat\Lambda(r)=I_r$. This normalization is used throughout the factor and diffusion-index analysis \citep{bai2008large,forni2000generalized,stock-watson-DFM}.

\begin{figure}[t!]
\centering
\includegraphics[width=.80\textwidth,keepaspectratio]{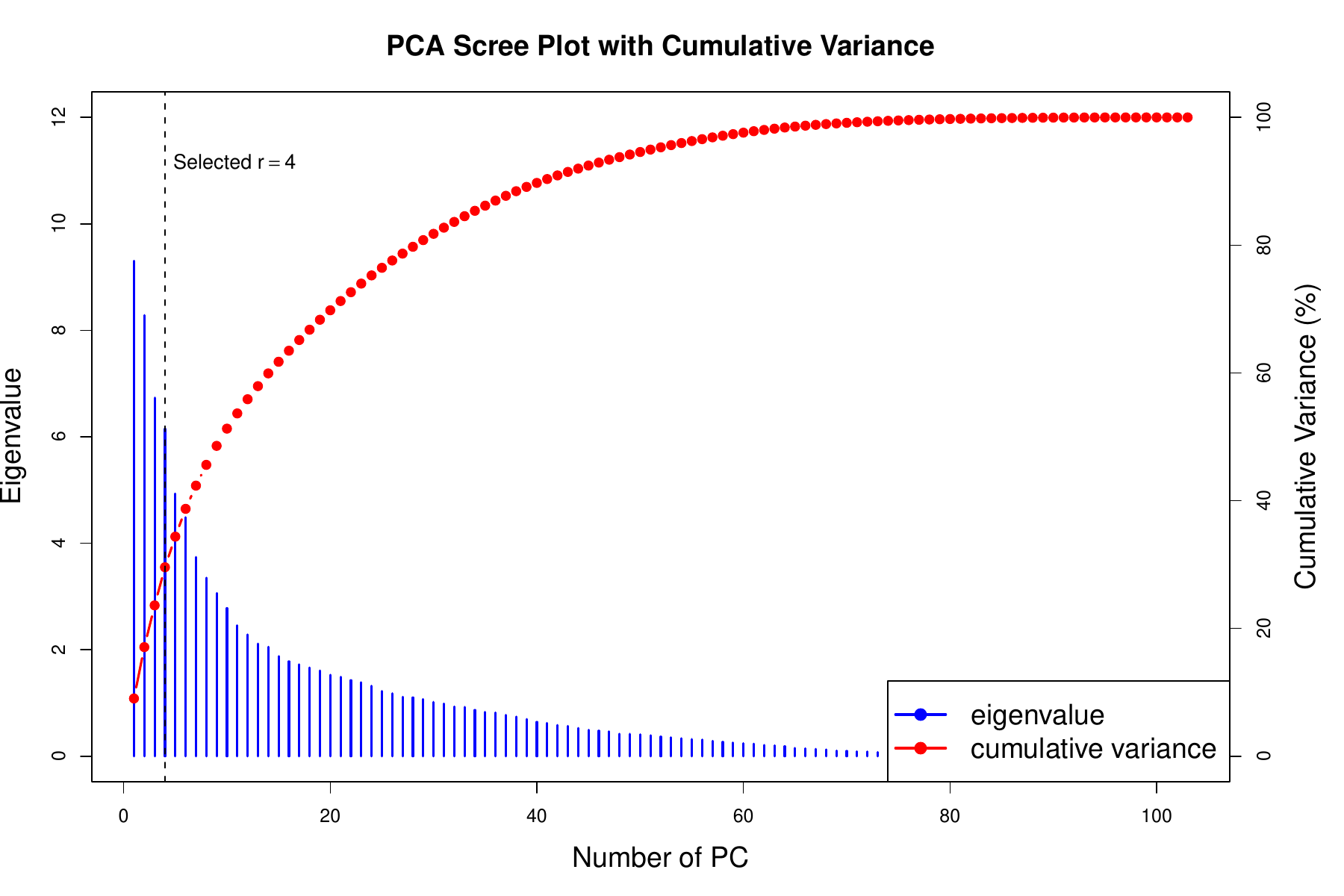}
\caption{Eigenvalue profile and cumulative variance share for the Korean static-factor panel.}
\label{f:screeplot}
\end{figure}

To determine the number of common factors, we rely on the eigenvalue-ratio (ER) and growth-ratio (GR) criteria of \citet{ahn2013eigenvalue}. Both criteria select four factors, and this choice is unchanged as the maximum candidate dimension varies from 8 to 30. The resulting four-dimensional principal-component space explains 29.58\% of the standardized panel variation, while the eigenvalue profile in Figure~\ref{f:screeplot} provides a complementary visual summary of the separation between the leading common components and the remaining variation.

The first two principal-component axes mix the same two economic blocks: industrial-production variables load with similar signs and broadly comparable magnitudes on both components, whereas interest-rate and term-spread variables load with opposite signs. For a transparent economic representation, let $\widehat z_{1t}$ and $\widehat z_{2t}$ denote the unit-variance versions of the first two principal-component scores and define the fixed $45^\circ$ rotation
\[
{
\widehat f_{1t}^{R}=\frac{\widehat z_{1t}-\widehat z_{2t}}{\sqrt{2}},
\qquad
\widehat f_{2t}^{R}=\frac{\widehat z_{1t}+\widehat z_{2t}}{\sqrt{2}}.}
\]

Before rotation, the otherwise arbitrary signs of $\widehat z_{1t}$ and $\widehat z_{2t}$ are normalized using \texttt{T3YFFM} and \texttt{IPMANSICS}, respectively; the resulting rotated factors are positively correlated with those same anchor series. The angle is fixed rather than estimated from auxiliary targets. The third and fourth factors retain the standardized third and fourth principal-component scores, with signs oriented toward \texttt{PERMIT} and \texttt{CE16OV}, respectively. This transformation preserves the first-two-factor span, the full four-factor space, and their cumulative explanatory power.\footnote{The first two scores are standardized before rotation because their eigenvalues differ; rotating the unstandardized scores would generally induce nonzero covariance between the resulting indices.} For an applied example of a fixed $45^\circ$ orthogonal rotation of the first two principal-component scores, see \citet{martel2021polychoric}.

\subsection{Interpreting the four factors}\label{s:factor-interpretation}

\begin{figure}[t!]
\centering
\includegraphics[width=.94\textwidth,keepaspectratio]{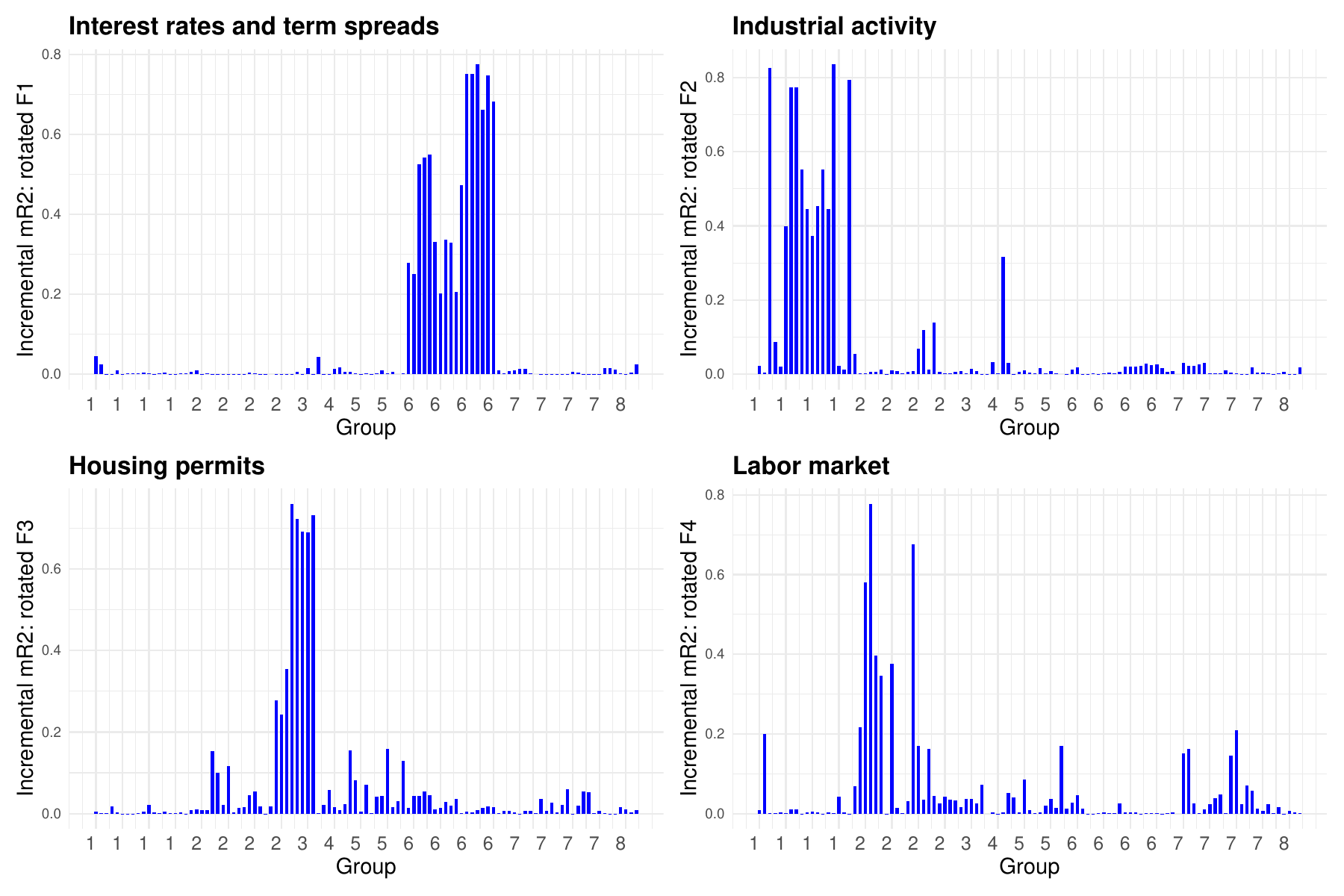}
\caption{Incremental explanatory power $mR_i^2(k)$ by KRED group for the fixed-$45^\circ$ factor representation, $k=1,\ldots,4$.}
\label{f:mR2_plot}
\end{figure}

To interpret the factors, we examine how much additional variation each one explains across the individual KRED series. For series $i$, let $R_i^2(k)$ denote the coefficient of determination from a regression on the first $k$ factors. The incremental contribution of factor $k$ is
\[
mR_i^2(1)=R_i^2(1),
\qquad
mR_i^2(k)=R_i^2(k)-R_i^2(k-1),\quad k=2,3,4.
\]
Figure~\ref{f:mR2_plot} displays these incremental contributions across the eight KRED groups, while Table~\ref{tab:mr2} lists the ten series most strongly associated with each factor. The fixed rotation cleanly separates the first two axes: the first is concentrated in interest rates and term spreads, whereas the second is concentrated in industrial production. Because the transformation is confined to the standardized first-two-component space, the cumulative fit after two and four factors is unchanged.

\begin{table}[t!]
\centering
\caption{Leading incremental $mR_i^2$ contributors in the four-factor specification.}
\footnotesize
\setlength{\tabcolsep}{4pt}
\begin{tabular}{lrc@{\hspace{1.4em}}lrc}
\toprule
\multicolumn{3}{c}{Factor 1} & \multicolumn{3}{c}{Factor 2}\\
\cmidrule(r){1-3}\cmidrule(l){4-6}
Series & $mR_i^2(1)$ & Group & Series & $mR_i^2(2)$ & Group\\
\midrule
\texttt{T3YFFM}      & 0.776 & 6 & \texttt{IPMANSICS}  & 0.836 & 1\\
\texttt{TB6SMFFM}    & 0.752 & 6 & \texttt{INDPRO}     & 0.827 & 1\\
\texttt{T1YFFM}      & 0.751 & 6 & \texttt{CUMFNS}     & 0.795 & 1\\
\texttt{AAAFFM}      & 0.747 & 6 & \texttt{IPCONGD}    & 0.774 & 1\\
\texttt{BAAFFM}      & 0.682 & 6 & \texttt{IPFINAL2}   & 0.773 & 1\\
\texttt{T10YFFM}     & 0.662 & 6 & \texttt{IPDCONGD}   & 0.552 & 1\\
\texttt{GS1}         & 0.550 & 6 & \texttt{IPDMAT}     & 0.552 & 1\\
\texttt{TB6MS}       & 0.543 & 6 & \texttt{IPMAT}      & 0.454 & 1\\
\texttt{TB3MS}       & 0.525 & 6 & \texttt{IPNCONGD}   & 0.446 & 1\\
\texttt{TB3SMFFM}    & 0.473 & 6 & \texttt{IPNMAT}     & 0.446 & 1\\
\midrule
\multicolumn{3}{c}{Factor 3} & \multicolumn{3}{c}{Factor 4}\\
\cmidrule(r){1-3}\cmidrule(l){4-6}
Series & $mR_i^2(3)$ & Group & Series & $mR_i^2(4)$ & Group\\
\midrule
\texttt{PERMIT}          & 0.760 & 3 & \texttt{CE16OV}       & 0.777 & 2\\
\texttt{PERMITW}         & 0.732 & 3 & \texttt{PAYEMS}       & 0.677 & 2\\
\texttt{PERMITNE}        & 0.722 & 3 & \texttt{CLF16OV}      & 0.581 & 2\\
\texttt{PERMITMW}        & 0.692 & 3 & \texttt{UNRATE}       & 0.397 & 2\\
\texttt{PERMITS}         & 0.691 & 3 & \texttt{UEMP5TO14}    & 0.376 & 2\\
\texttt{CES3000000008}   & 0.354 & 2 & \texttt{UEMPMEAN}     & 0.346 & 2\\
\texttt{CES0600000008}   & 0.278 & 2 & \texttt{HWIURATIO}    & 0.217 & 2\\
\texttt{CES2000000008}   & 0.244 & 2 & \texttt{CUSR0000SAC}  & 0.209 & 7\\
\texttt{NONREVSL}        & 0.159 & 5 & \texttt{GNI\_real}    & 0.200 & 1\\
\texttt{M1SL}            & 0.156 & 5 & \texttt{DTCTHFNM}     & 0.170 & 5\\
\bottomrule
\end{tabular}
\label{tab:mr2}
\end{table}

\noindent\textbf{Factor 1: interest rates and term spreads.}\;
The first rotated factor is concentrated in the interest-rate block. Its ten largest incremental contributions all come from short- and medium-term yields or spreads relative to the policy rate, led by \texttt{T3YFFM}, \texttt{TB6SMFFM}, \texttt{T1YFFM}, and \texttt{AAAFFM}. It therefore summarizes the common movement in rates, the yield curve, and risk premia \citep{bernanke1995inside,gurkaynak2005sensitivity}. Because a wider term spread can reflect policy expectations, growth prospects, or risk compensation, the factor is not interpreted mechanically as a one-directional index of financial easing or tightening.

\smallskip
\noindent\textbf{Factor 2: industrial activity.}\;
The second rotated factor is concentrated in manufacturing output, goods production, and capacity utilization. \texttt{IPMANSICS}, \texttt{INDPRO}, \texttt{CUMFNS}, \texttt{IPCONGD}, and \texttt{IPFINAL2} provide the largest incremental contributions, while interest-rate variables make only small contributions. This pattern identifies a coincident industrial cycle of the type emphasized in the business-cycle literature \citep{stock1989new} and separates it from the interest-rate and term-spread axis captured by Factor~1.

\smallskip
\noindent\textbf{Factor 3: housing permits.}\;
The third factor is dominated by the national and regional housing-permit series, which occupy the first five positions in the incremental-$R^2$ ranking. Because permits are forward-looking indicators of construction and residential investment \citep{strauss2013does}, the factor captures changes in the housing pipeline before they are fully reflected in completed activity. Wage and credit variables enter farther down the ranking and provide secondary links rather than defining the factor.

\smallskip
\noindent\textbf{Factor 4: labor market.}\;
The fourth factor is anchored by employment, labor-force participation, unemployment, and unemployment-duration measures. It therefore summarizes labor utilization and slack, while the smaller contributions from selected price and credit variables reflect the broader macroeconomic environment in which labor-market adjustment takes place.

\begin{figure}[t!]
\centering
\includegraphics[width=.90\textwidth,keepaspectratio]{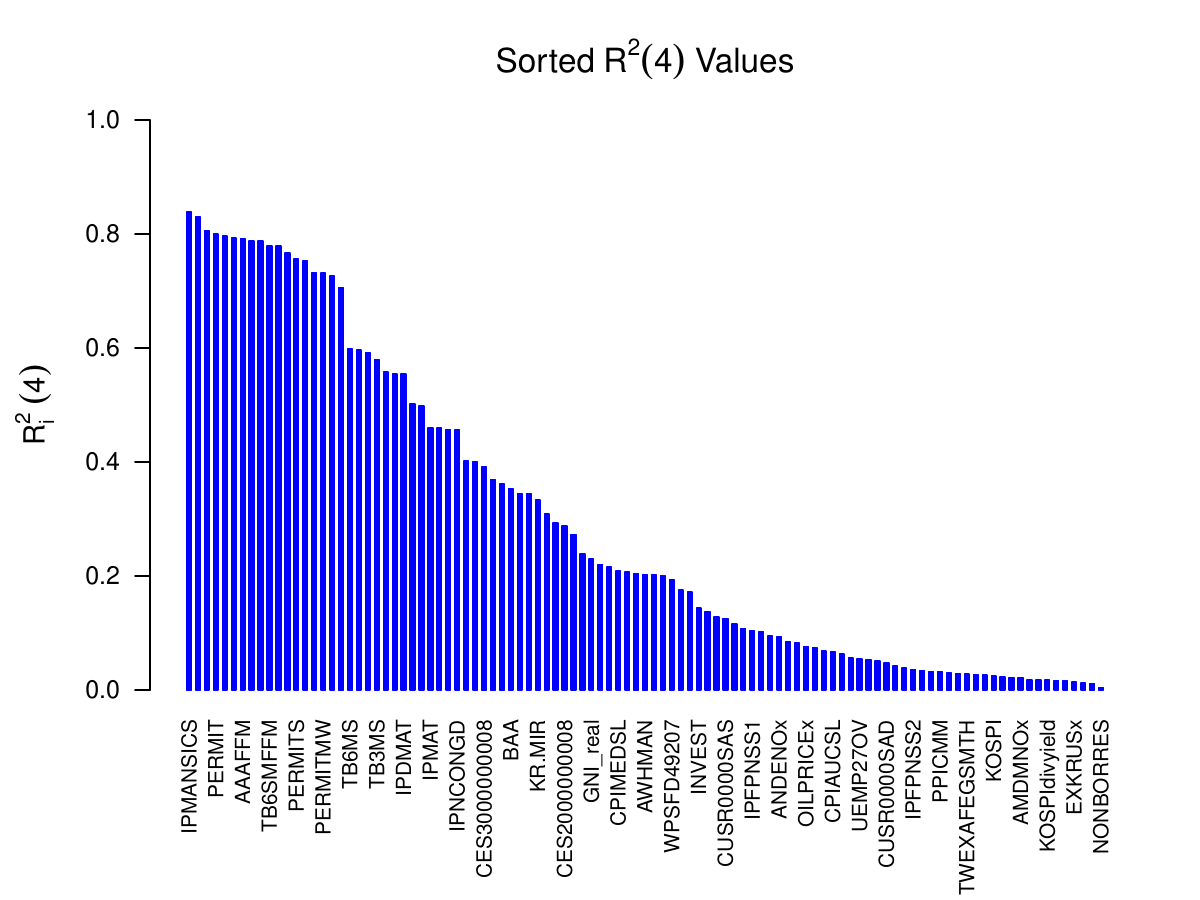}
\caption{Series-level fit from the four-factor specification.}
\label{f:barplot_r4}
\end{figure}

Figure~\ref{f:barplot_r4} ranks the series-level fit $R_i^2(4)$ after all four factors are included. The best-explained variables come from four connected blocks---yield-curve and credit-spread measures, manufacturing production and utilization, housing permits, and labor-market quantities---showing that the factor structure captures a broad macro-financial core of the Korean economy. More sector-specific series appear farther down the ranking and remain largely in the idiosyncratic component.

\subsection{Factor-based diffusion indices}\label{s:diffusion}

Following \citet{mccracken2015fredmd}, we construct factor-based diffusion indices to summarize low-frequency dynamics implied by the estimated common components. Rather than tracking the share of series increasing at each date, the factor-based index cumulates each estimated factor score:
$$
\widehat{\mathrm{FDI}}_{t}(k)=\sum_{s=1}^{t}\widehat f_{s}(k),\qquad t=1,\ldots,T,
$$
where $\widehat f_{s}(k)$ denotes the $k$-th score in the fixed-$45^\circ$ four-factor representation defined in Section~\ref{s:factor-number}. This cumulation accentuates persistent expansions and contractions.

\begin{figure}[t!]
	\centering
	\includegraphics[width=.94\textwidth,keepaspectratio]{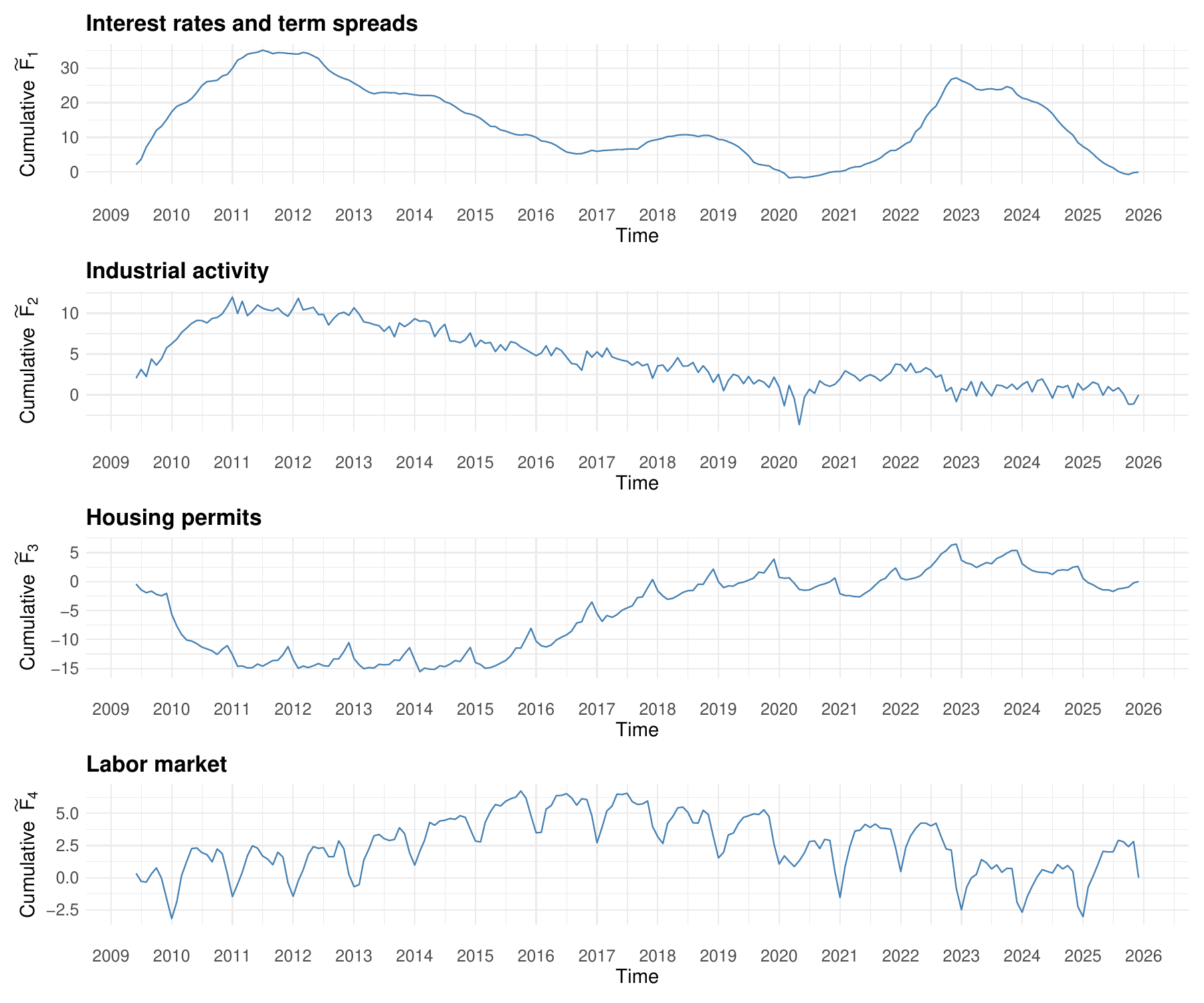}
	\caption{Factor-based diffusion indices for the fixed-$45^\circ$ four-factor representation.}
	\label{f:diffusion}
\end{figure}

Figure~\ref{f:diffusion} plots the four diffusion indices and highlights several episodes.
The interest-rate-and-term-spread index ($F_1$) rises in the aftermath of the global financial crisis, when the yield curve was steep and term spreads remained elevated, and then declines during the prolonged low-interest-rate period of the 2010s. It remains subdued around the pandemic as policy and market rates fall, before increasing sharply in 2021--2022 with monetary-policy normalization and bond-market repricing. The index then declines through 2023--2025 as the yield curve flattens and term spreads narrow. The industrial-activity index ($F_2$) rises during the early post-crisis recovery, erodes gradually over 2012--2019, and records the sharp contraction and rebound associated with the pandemic. It strengthens during 2021, weakens markedly in 2022, stabilizes partially in 2023--2024, and softens during much of 2025, followed by a late rebound. Housing-permit diffusion ($F_3$) is weak through the first half of the sample but trends upward from about 2016, peaks around the end of 2022, and then eases. The labor-market index ($F_4$) rises gradually through the mid-2010s, falls during the pandemic, partially recovers in 2021--2022, and remains volatile and below its earlier peak thereafter.

The pandemic and subsequent normalization therefore produce a clear divergence across the rotated factors. Industrial activity exhibits a sharp within-2020 contraction and reopening rebound, while the interest-rate and term-spread component remains subdued until its strong 2021--2022 rise. Housing and labor-market adjustment is smoother and more delayed. After 2022, the rate-and-spread index falls, industrial activity remains uneven, housing turns down, and labor-market diffusion stays below its earlier peak despite a partial rebound.

These patterns suggest a sequencing in which interest rates and term spreads respond sharply to policy normalization and market repricing, industrial activity exhibits episodic real-side swings, housing follows a slower medium-run cycle, and labor-market conditions adjust most gradually. The four-factor representation therefore summarizes major Korean macroeconomic episodes while separating interest-rate, industrial, housing, and labor-market forces.

\section{FAVAR evidence on monetary policy shocks} \label{s:favar}

We next use KRED in a factor-augmented vector autoregression to study how U.S. and Korean monetary-policy shocks propagate through a broad domestic information set. The analysis follows the data-rich identification framework of \citet{bernanke2005measuring}, hereafter BBE, allowing the estimated common factors to summarize information omitted from a standard VAR.

\subsection{BBE identification and lag selection}\label{s:favar-identification}

{
Following \citet{bernanke2005measuring}, we classify the information set into 56 slow-moving and 46 fast-moving variables and treat \texttt{US.FFR} and \texttt{KR.MIR} as observed policy rates.\footnote{The slow/fast classification follows the BBE within-month timing principle. Slow-moving variables consist primarily of real-activity, labor-market, realized-consumption, and official price measures, whereas housing, orders and sentiment, money and credit, market interest rates and spreads, exchange rates, commodity prices, and equity-market variables are treated as fast-moving. Housing starts and permits are also treated as fast-moving, consistent with the classification in BBE. The two policy rates are classified separately as observed variables.}
With $K=4$, let $C_t$ denote the first $K+2=6$ principal components of the full panel and $F_t^s$ the first $K=4$ principal components of the slow-moving block. We estimate
}
\[
{
C_t=a+A F_t^s+B Y_t+e_t,\qquad
Y_t=(\texttt{US.FFR}_t,\texttt{KR.MIR}_t)' ,
}
\]
{
and remove the contemporaneous policy-rate component as $\widetilde C_t=C_t-\widehat B Y_t$. The cleaned components are restandardized and compressed to $K=4$ final factors, which enter the VAR together with the two observed policy rates. These FAVAR factors are constructed separately from the descriptive rotated factors in Section~\ref{s:factor-number}.}
The resulting state vector is
\[
y_t=
\bigl(\texttt{US.FFR}_t,\widehat F_{1t},\widehat F_{2t},
\widehat F_{3t},\widehat F_{4t},\texttt{KR.MIR}_t\bigr)'.
\]
Structural innovations are identified recursively in the same order. Ordering \texttt{US.FFR} first imposes the small-open-economy restriction that Korean variables do not contemporaneously affect U.S. policy at a monthly frequency; ordering \texttt{KR.MIR} last allows Korean policy to react within the month to the external and latent information block.

BIC and HQ select two lags for both the FAVAR and the four-variable benchmark VAR. We therefore use the two-lag system $(\texttt{US.FFR}_t,\widehat F_{1t},\ldots,\widehat F_{4t},\texttt{KR.MIR}_t)'$ as the preferred BBE-FAVAR(2). The corresponding low-dimensional benchmark is the VAR(2) in $(\texttt{US.FFR}_t,\texttt{IP}_t,\texttt{CPI}_t,\texttt{KR.MIR}_t)'$, where \texttt{IP} and \texttt{CPI} denote the Korean industrial-production and all-items consumer-price series, respectively. KRED retains the FRED-MD-style mnemonics \texttt{INDPRO} and \texttt{CPIAUCSL} for these Korean counterparts. To assess sensitivity to lag length, we also estimate the BBE-FAVAR(13) and VAR(13).

\subsection{Bootstrap inference}\label{s:favar-bootstrap}

We construct confidence intervals using Procedure A of \citet{yamamoto2019bootstrap}. In a FAVAR, the common factors are estimated from a large panel rather than directly observed. Procedure A therefore re-estimates both the factors and the FAVAR in each bootstrap sample, so the confidence intervals reflect uncertainty from both factor extraction and VAR estimation. The main analysis uses 3000 bootstrap replications.

Following the finite-sample implementation in \citet{yamamoto2019bootstrap}, we apply a Kilian-style small-sample bias correction to the VAR coefficients \citep{kilian1998small}. Our primary 90\% confidence intervals are Hall percentile intervals constructed from recentered bootstrap impulse-response errors (hereafter, Hall-recentered intervals) \citep{hall1992bootstrap}. We also report Efron percentile intervals, which use the empirical quantiles of the bootstrap impulse responses directly, as a sensitivity check \citep{efron1993bootstrap}. These methods address distinct parts of the inference problem: Procedure A accounts for factor-estimation uncertainty, the Kilian correction reduces small-sample bias in the estimated VAR dynamics, and the Hall and Efron methods provide alternative ways of constructing confidence intervals.

Residuals in a large macroeconomic panel may also be related across series and over time. We therefore supplement the baseline Procedure A bootstrap with dependent wild bootstrap (DWB) calculations based on \citet{shao2010dependent} and block wild bootstrap (BWB) calculations motivated by
\citet{djogbenou2015bootstrap}. For background on wild-bootstrap inference in autoregressive models, see \citet{goncalves2004bootstrapping}. These alternatives examine whether the economic conclusions are robust to residual dependence. For the DWB, $L$ denotes the bandwidth governing the temporal dependence among the wild multipliers, whereas for the BWB, it denotes the block length. In both cases, a larger value of $L$ accommodates dependence over a longer horizon. Given the monthly frequency of the data, we consider $L=6$ and $L=12$, corresponding to dependence windows of approximately six and twelve months, respectively. Detailed implementation, residual diagnostics, and numerical-stability results are reported in Appendix~\ref{apdx:favar-robustness}.

\subsection{U.S. and Korean monetary-policy shocks}\label{s:favar-irf}

The reported responses cover the two policy rates, industrial production, CPI, the Denton-based monthly real GDP and real GNI series described in Section~\ref{s:data-preprocessing}, unemployment, real M2, the one-year and three-year term spreads, the won--dollar exchange rate, and the KOSPI. For brevity, we refer to the first of the two Denton-based series as monthly real GDP below.

\begin{figure}[p]
\centering
\includegraphics[width=.94\textwidth,height=.78\textheight,keepaspectratio]{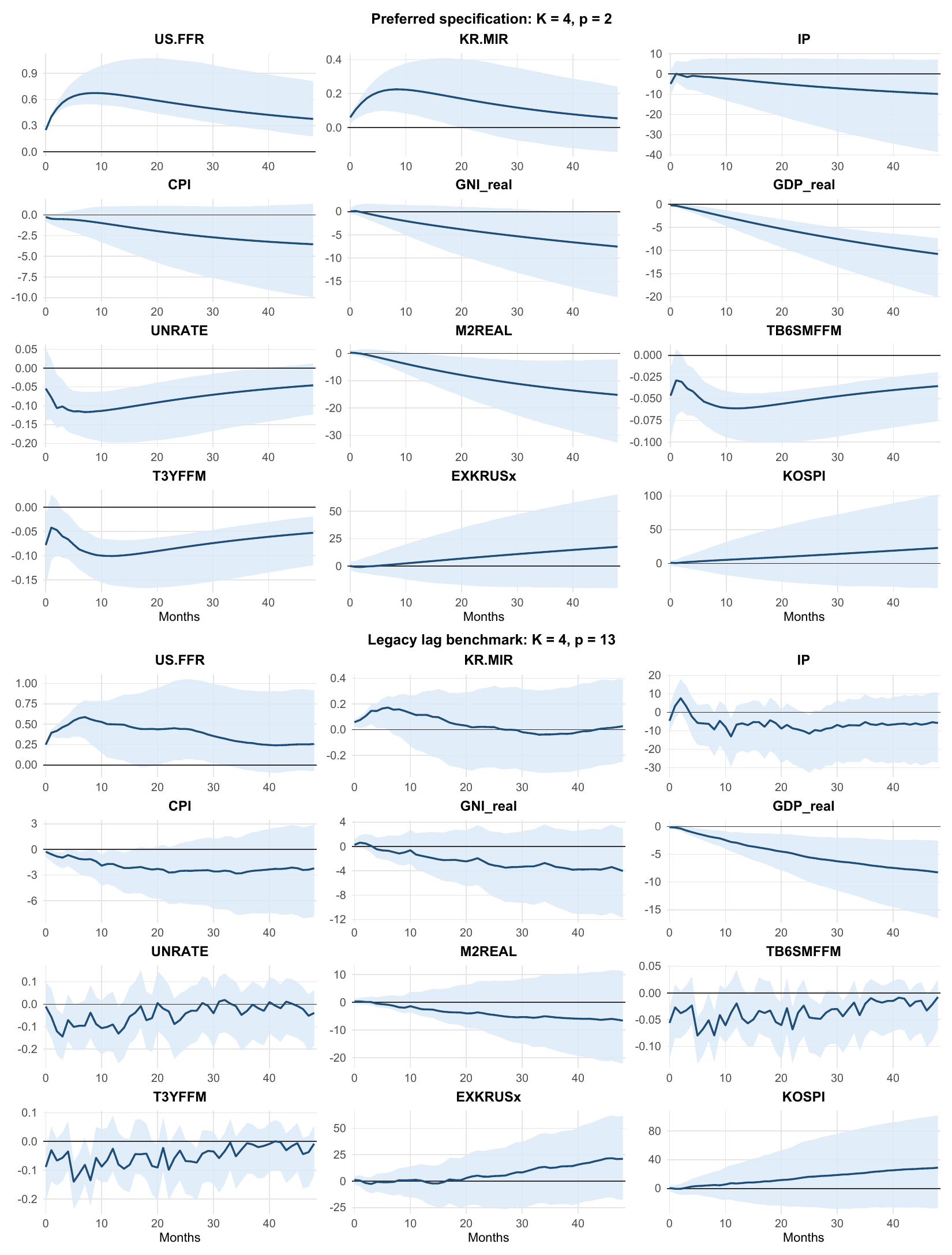}
\caption{Responses to a 25-basis-point U.S. policy-rate shock. The figure compares the preferred BBE-FAVAR(2) with the BBE-FAVAR(13) sensitivity specification; shaded regions are 90\% bootstrap confidence intervals. Interest rates, unemployment, and term spreads are reported in percentage-point units. Log-based level responses are multiplied by $10^3$ (one plotted unit is approximately a 0.1\% level change). Since \texttt{CPIAUCSL} enters as $\Delta\log CPI$, its response is cumulated before plotting and therefore represents the CPI price level.}
\label{f:favar-USimpact}
\end{figure}

Figure~\ref{f:favar-USimpact} reports responses to a 25-basis-point U.S. policy-rate shock under the preferred two-lag model and the conventional 13-lag sensitivity benchmark. In the preferred specification, monthly real GDP declines persistently and \texttt{KR.MIR} rises, indicating a domestic policy response to U.S. tightening. The CPI point response is generally nonpositive, although its 90\% confidence intervals include zero; as in BBE, the KRED-based FAVAR substantially attenuates the price puzzle relative to the small VAR. We examine this comparison in detail in the next subsection. The responses of industrial production and the KOSPI are also imprecisely estimated, so the strongest conclusions concern monthly real GDP and the Korean policy rate.

\begin{figure}[p]
\centering
\includegraphics[width=.94\textwidth,height=.78\textheight,keepaspectratio]{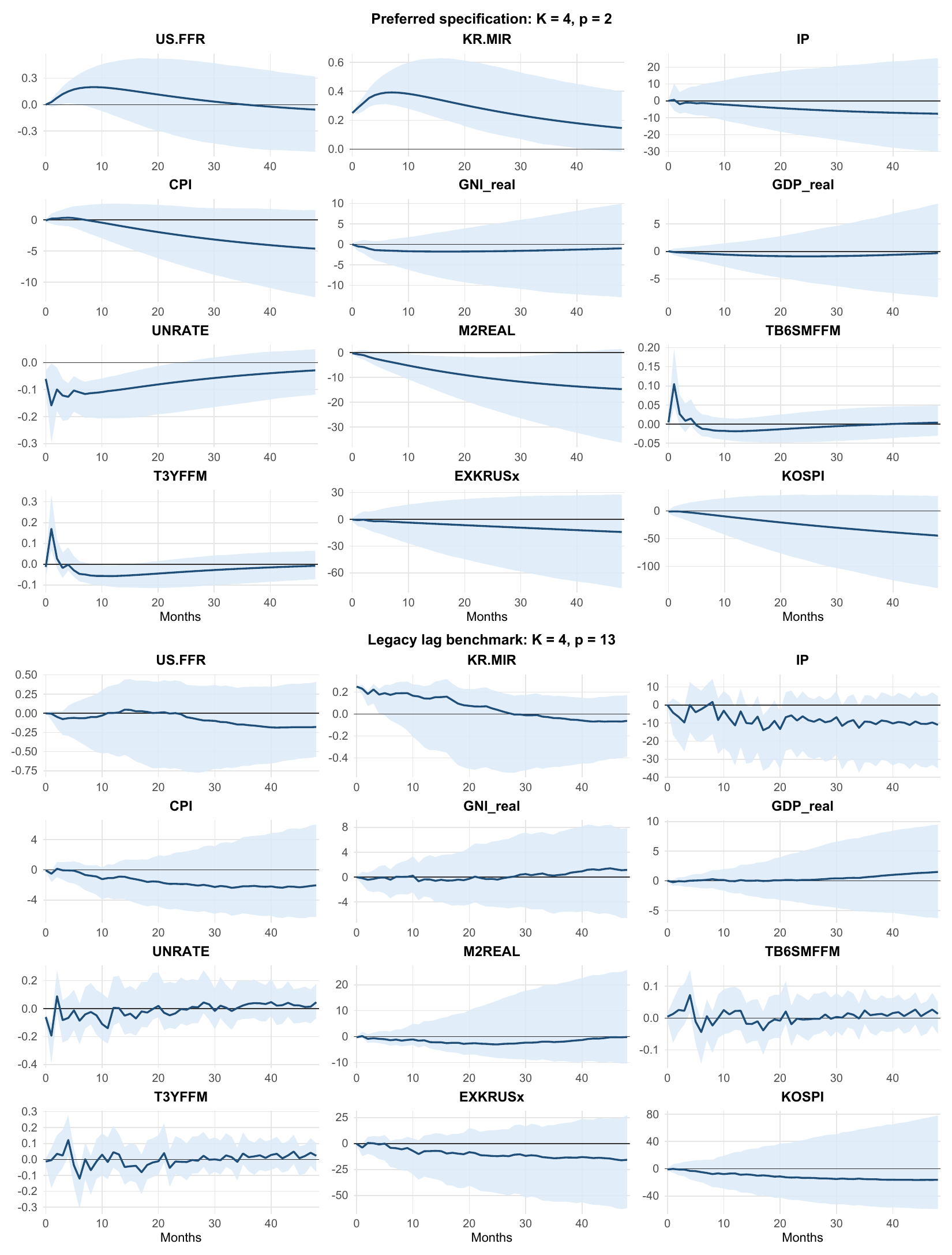}
\caption{Responses to a 25-basis-point Korean policy-rate shock. The figure compares the preferred BBE-FAVAR(2) with the BBE-FAVAR(13) sensitivity specification; shaded regions are 90\% bootstrap confidence intervals. Interest rates, unemployment, and term spreads are reported in percentage-point units. Log-based level responses are multiplied by $10^3$ (one plotted unit is approximately a 0.1\% level change). Since \texttt{CPIAUCSL} enters as $\Delta\log CPI$, its response is cumulated before plotting and therefore represents the CPI price level.}
\label{f:favar-KRimpact}
\end{figure}

Figure~\ref{f:favar-KRimpact} reports responses to a 25-basis-point Korean policy-rate shock. The domestic policy rate rises on impact and subsequently mean-reverts, while the broader panel reveals heterogeneous responses across activity, prices, credit, exchange rates, and equity markets. Although the point responses of many variables---including industrial production, GDP, M2, the won--dollar exchange rate, and the KOSPI---broadly follow conventional monetary-transmission patterns, the confidence intervals remain wide. The CPI price-level response is less clear at short horizons. The figure nevertheless illustrates the broad range of monetary-policy transmission channels that can be examined using the KRED information set.

\subsection{Price-puzzle comparison and the 13-lag benchmark}\label{s:price-puzzle}

\begin{figure}[t!]
	\centering
	\includegraphics[width=.94\textwidth,keepaspectratio]{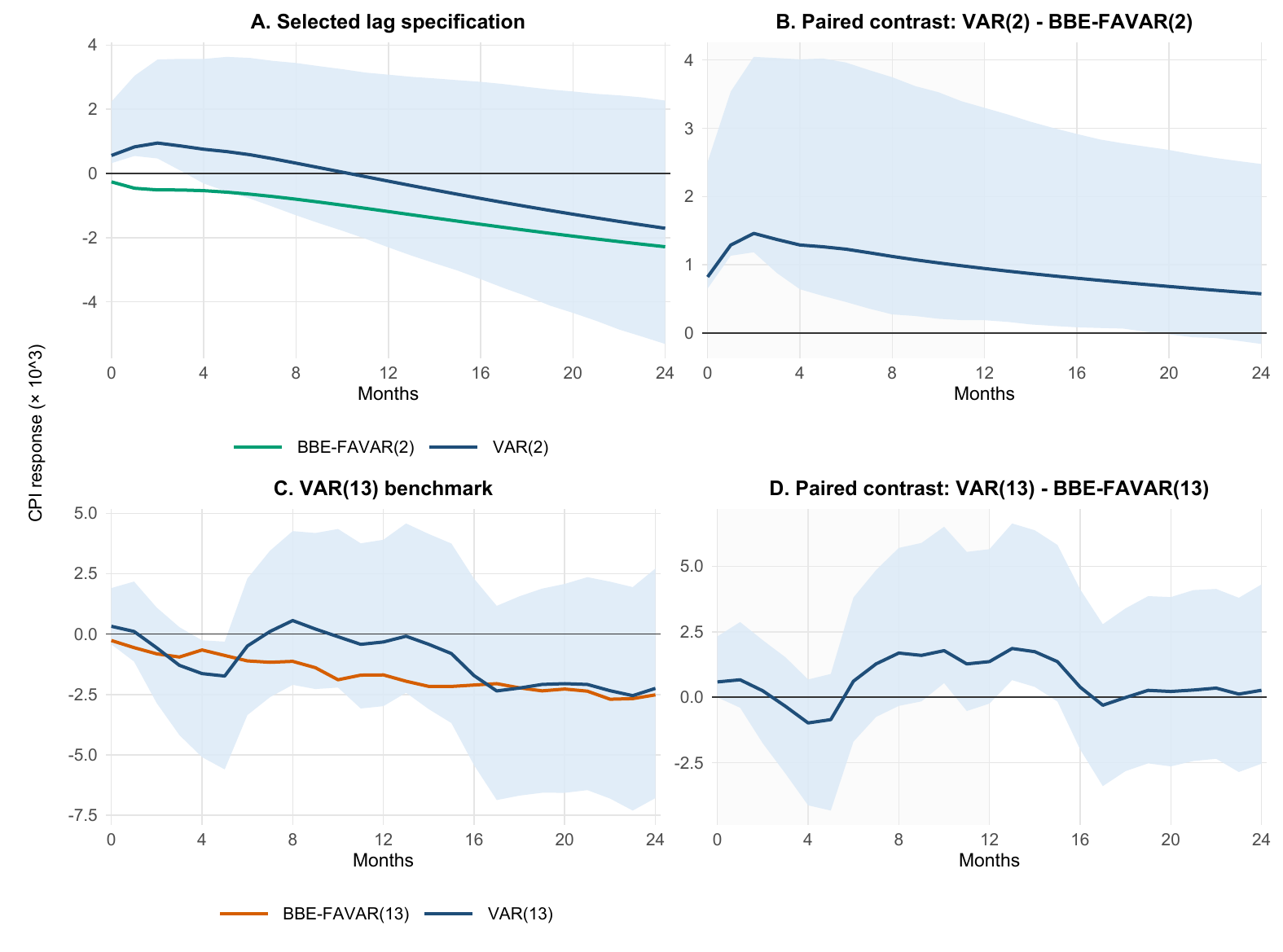}
	\caption{CPI price-level responses and paired VAR--BBE-FAVAR differences. Panels A--B report the preferred two-lag models; Panels C--D report the 13-lag sensitivity models. Responses are $10^3$ times the cumulative $\Delta\log CPI$ response, so one plotted unit is approximately a 0.1\% change in the CPI level.}
	\label{f:price-puzzle-comparison}
\end{figure}

The preferred two-lag specification provides clear evidence that the broader KRED information set attenuates the price puzzle generated by a small VAR. We summarize the short-run CPI price-level responses over horizons 0--12 using the mean VAR--FAVAR contrast, the reduction in the maximum positive response, and the reduction in positive response area. Figure~\ref{f:price-puzzle-comparison} displays the complete response paths and paired differences through month 24.

As Table~\ref{tab:price-puzzle-points} shows, the small VAR produces a positive mean CPI price-level response, with positive responses at 11 of the first 13 horizons. By contrast, the BBE-FAVAR price-level response is nonpositive throughout, with a negative mean response. The resulting mean contrast of 0.001159 and peak-response reduction of 0.001211 indicate an economically sizable reduction in both the average and the largest anomalous price-level response.

\begin{table}[t!]
\centering
\caption{CPI price-level response comparison over horizons 0--12. Entries are in cumulative log units; the corresponding response paths in Figure~\ref{f:price-puzzle-comparison} are multiplied by $10^3$.}
\small
\begin{tabular}{lrrrrrr}
\toprule
Lag & VAR mean & FAVAR mean & Mean contrast & Peak reduction & VAR $>0$ & FAVAR $>0$\\
\midrule
$p=2$  & 0.000454 & $-0.000706$ & 0.001159 & 0.001211 & 11/13 & 0/13\\
$p=13$ & $-0.000405$ & $-0.001093$ & 0.000687 & 0.000826 & 5/13 & 0/13\\
\bottomrule
\end{tabular}
\label{tab:price-puzzle-points}
\end{table}

The statistical evidence reinforces this conclusion. Table~\ref{tab:price-puzzle-hall} shows that the primary 90\% Hall-recentered intervals exclude zero for both the mean contrast and the peak-response reduction. The same conclusion holds under every DWB and BWB design, indicating that the attenuation is not driven by treating the panel residuals as independent.

\begin{table}[t!]
\centering
\caption{Primary 90\% bootstrap intervals for paired CPI price-level attenuation, $p=2$. Interval endpoints are reported in cumulative log units.}
\small
\begin{tabular}{lcc}
\toprule
Bootstrap design & Mean VAR--FAVAR contrast & Peak-response reduction\\
\midrule
Procedure A & $[0.000582,\ 0.003605]$ & $[0.000665,\ 0.003313]$\\
DWB, $L=6$  & $[0.000707,\ 0.003665]$ & $[0.000878,\ 0.003422]$\\
DWB, $L=12$ & $[0.000647,\ 0.003701]$ & $[0.000765,\ 0.003377]$\\
BWB, $L=6$  & $[0.000515,\ 0.003665]$ & $[0.000567,\ 0.003461]$\\
BWB, $L=12$ & $[0.000800,\ 0.003651]$ & $[0.000910,\ 0.003376]$\\
\bottomrule
\end{tabular}
\label{tab:price-puzzle-hall}
\end{table}

The alternative Efron percentile intervals include zero, and the reduction in total positive response area is not statistically significant. These supplementary results do not overturn the main finding but limit its scope: the evidence establishes a robust attenuation of the anomalous positive response in the two-lag small VAR benchmark, rather than a precisely estimated fall in the price level itself.

The 13-lag models provide a useful contrast. In that specification, the small VAR already has a negative average CPI price-level response and the paired differences are not statistically significant, so the price-puzzle result is specific to the empirically preferred two-lag dynamics. We therefore treat the 13-lag models as a lag-length sensitivity exercise. Appendix Figure~\ref{f:appendix-price-puzzle-bootstrap} reports the corresponding dependence-robust comparisons.

\section{U.S.--Korea dependence based on tensor autoregression} \label{s:tenar}

This section uses the common grouped architecture of KRED and FRED-MD to study where U.S.--Korea dependence is concentrated across macroeconomic blocks. The motivating question is economic rather than purely methodological. Korea is a small open economy, and the preceding FAVAR results indicate that external monetary conditions, especially U.S. tightening, matter for Korean macroeconomic dynamics. The remaining question is whether this dependence is broad-based across real and financial sectors or concentrated in a narrower set of transmission channels.

To address this question, we construct grouped monthly factors for Korea and the United States using a common eight-group classification. This grouped design allows us to compare the two countries within a unified framework and to distinguish broad real-side comovement from sharper macro-financial spillovers. In particular, it allows us to assess whether cross-country dependence is strongest in real activity, housing, money and credit, financial conditions, or equity-market blocks. The analysis proceeds in two stages. The coefficient matrices first describe lag-specific connections across country--group blocks, and the impulse responses then assess whether those connections imply directional propagation.

Let $X_t \in \mathbb{R}^{2 \times 8}$ denote the monthly grouped-factor matrix, where the two rows correspond to Korea and the United States and the eight columns correspond to the common macroeconomic groups. We model $\{X_t\}$ by a matrix-valued tensor autoregression, which is a second-order special case of the tensor autoregressive models studied by \citet{chen2021matrixar,hill2021tensor}:
\begin{equation}
    X_t = \sum_{i=1}^P \sum_{r=1}^{R_i} A_c^{(ir)} X_{t-i} A_g^{(ir)\prime} + E_t,
    \label{eq:tenar-main}
\end{equation}
where $A_c^{(ir)} \in \mathbb{R}^{2 \times 2}$ governs dependence across countries, $A_g^{(ir)} \in \mathbb{R}^{8 \times 8}$ governs dependence across groups, and $E_t$ is a mean-zero innovation matrix. For stability analysis and impulse-response calculations, we use the vectorized representation
\begin{equation}
    \mathrm{vec}(X_t) = \sum_{i=1}^P \Phi_i \mathrm{vec}(X_{t-i}) + \varepsilon_t,
    \qquad
    \Phi_i = \sum_{r=1}^{R_i} A_g^{(ir)} \otimes A_c^{(ir)},
    \label{eq:tenar-vec}
\end{equation}
with $\varepsilon_t = \mathrm{vec}(E_t)$.

The sample runs from 2009:06 to 2025:12. This application uses a country--group factor construction that is separate from the aggregate Korean static-factor panel and FAVAR information set used in Sections~\ref{s:empirical-analysis} and~\ref{s:favar}. Quarterly series and sporadic missing observations are adjusted at the level stage before the transformation codes are applied, and no post-transformation interpolation is used. For each country $c \in \{\mathrm{KR}, \mathrm{US}\}$ and each group $g \in \{1,\ldots,8\}$, we extract the first principal component $f_{cgt}$ from all transformed series in that block, yielding
\[
X_t =
\begin{pmatrix}
    f_{\mathrm{KR},1,t} & \cdots & f_{\mathrm{KR},8,t} \\
    f_{\mathrm{US},1,t} & \cdots & f_{\mathrm{US},8,t}
\end{pmatrix},
\qquad t = 1,\ldots,T.
\]
This construction does not impose exact one-to-one matching across the two national panels. Instead, it uses all available series within each country--group block so that the common grouped architecture is preserved.

We estimate \eqref{eq:tenar-main} by least squares using the \texttt{tensorTS} package \citep{chen2025tensorts}, and use maximum likelihood as a robustness check. Tuning parameters are selected by BIC subject to stability of the companion form. We search over $P \in \{1,2,3,6,12,13\}$, with $R_i \equiv R \in \{1,2\}$ for $P \le 3$ and $R_i \equiv 1$ for $P \in \{6,12,13\}$. The preferred specification has $P=2$ and $R=2$, with maximum companion-root modulus 0.958.

Two sign normalizations are important for interpretation. The financial factor $G6$ is oriented so that larger values correspond to higher rates and wider spreads, so a positive shock represents tighter financial conditions. The stock-market factor $G8$ is oriented so that larger values correspond to stronger equity prices, so a negative response represents stock-market weakness.

\begin{figure}[t!]
	\centering
	\includegraphics[width=.92\textwidth,keepaspectratio]{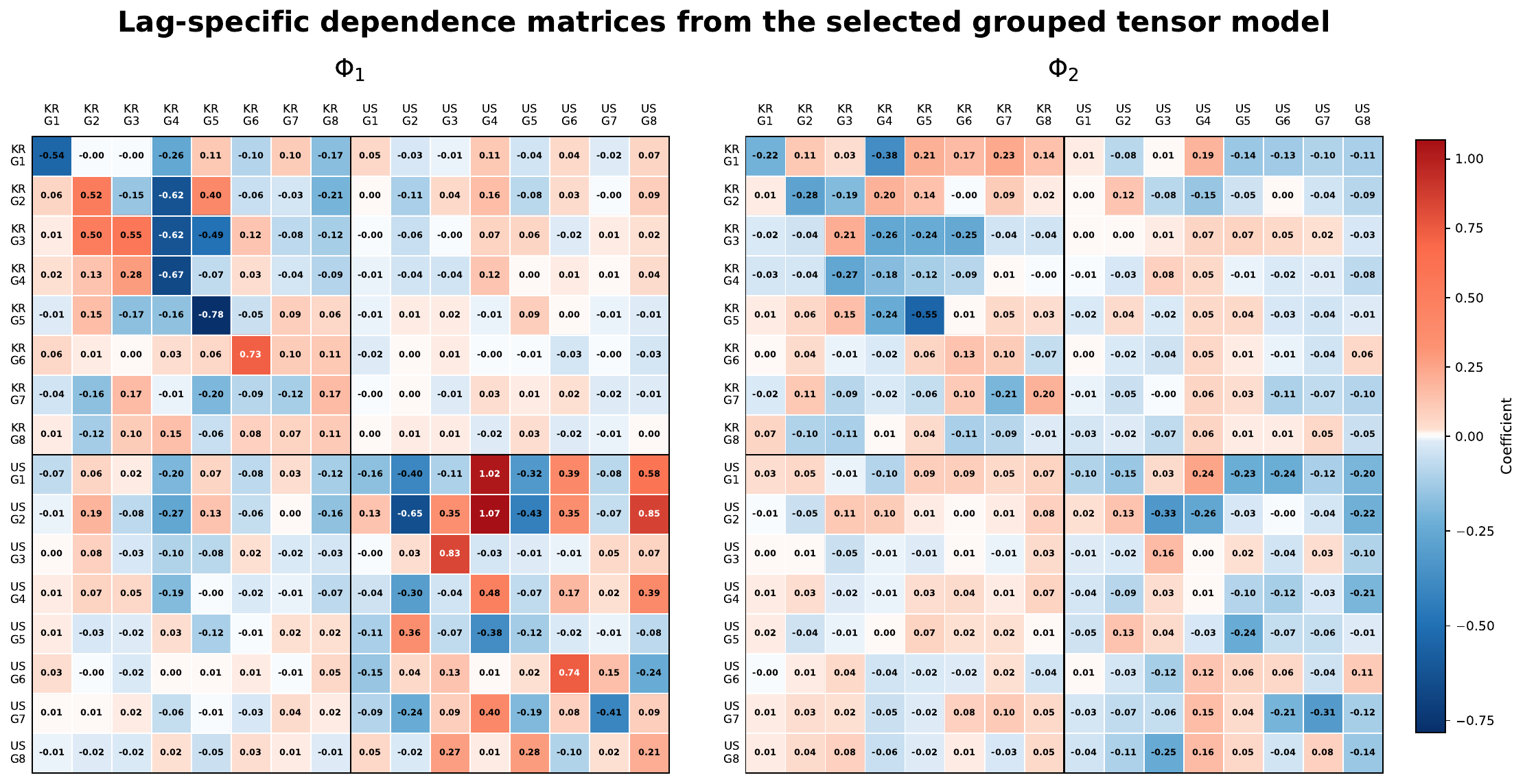}
	\caption{Lag-specific dependence matrices $\widehat\Phi_1$ and $\widehat\Phi_2$ implied by the BIC-selected grouped tensor model. More saturated cells indicate coefficients with larger absolute magnitudes, and the cell labels report the estimated values.}
	\label{f:tenar-dependence-matrix}
\end{figure}

Figure~\ref{f:tenar-dependence-matrix} reports the lag-specific coefficient matrices $\widehat\Phi_1$ and $\widehat\Phi_2$ implied by the BIC-selected grouped tensor model. Two features stand out. First, the main dependence structure is concentrated in $\widehat\Phi_1$, while $\widehat\Phi_2$ is weaker and more diffuse. Second, the largest coefficients are found within country blocks, so the dominant pattern remains one of strong domestic persistence rather than broad cross-country coupling. Within this overall structure, the consumption, orders, and inventories block ($G4$) is notable. In both countries, $G4$ is visibly connected to the real-side blocks $G1$--$G4$, suggesting that this block acts as a broad real-side adjustment margin linking output, labor, and demand-related fluctuations. Some cross-country entries involving $G4$ are also more visible than those of several other nonfinancial blocks. This pattern is consistent with an international demand--orders--inventories channel, although the coefficient matrices alone do not indicate a strongly directional spillover.

\begin{figure}[t!]
	\centering
	\includegraphics[width=.78\textwidth,keepaspectratio]{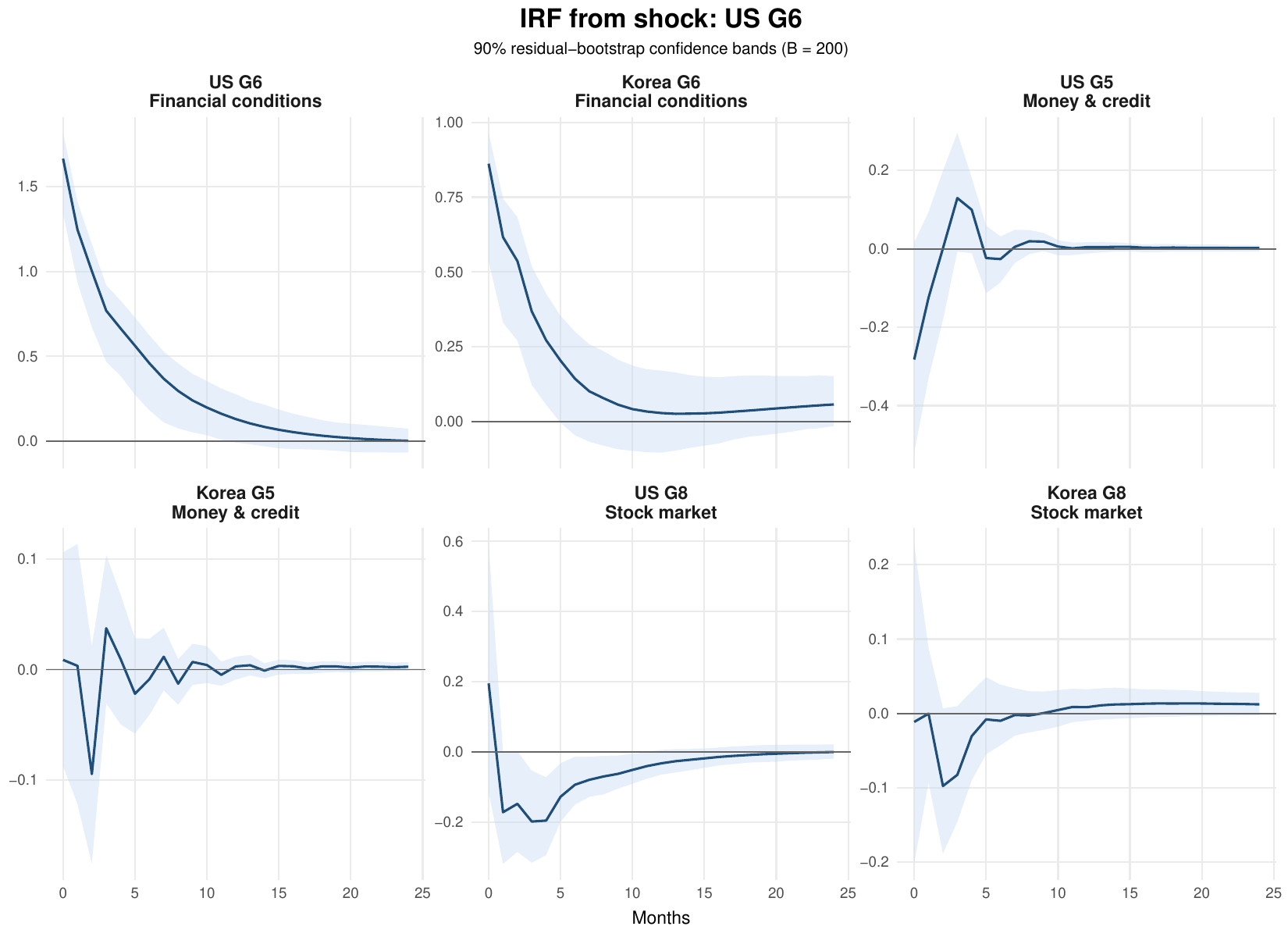}
	\includegraphics[width=.78\textwidth,keepaspectratio]{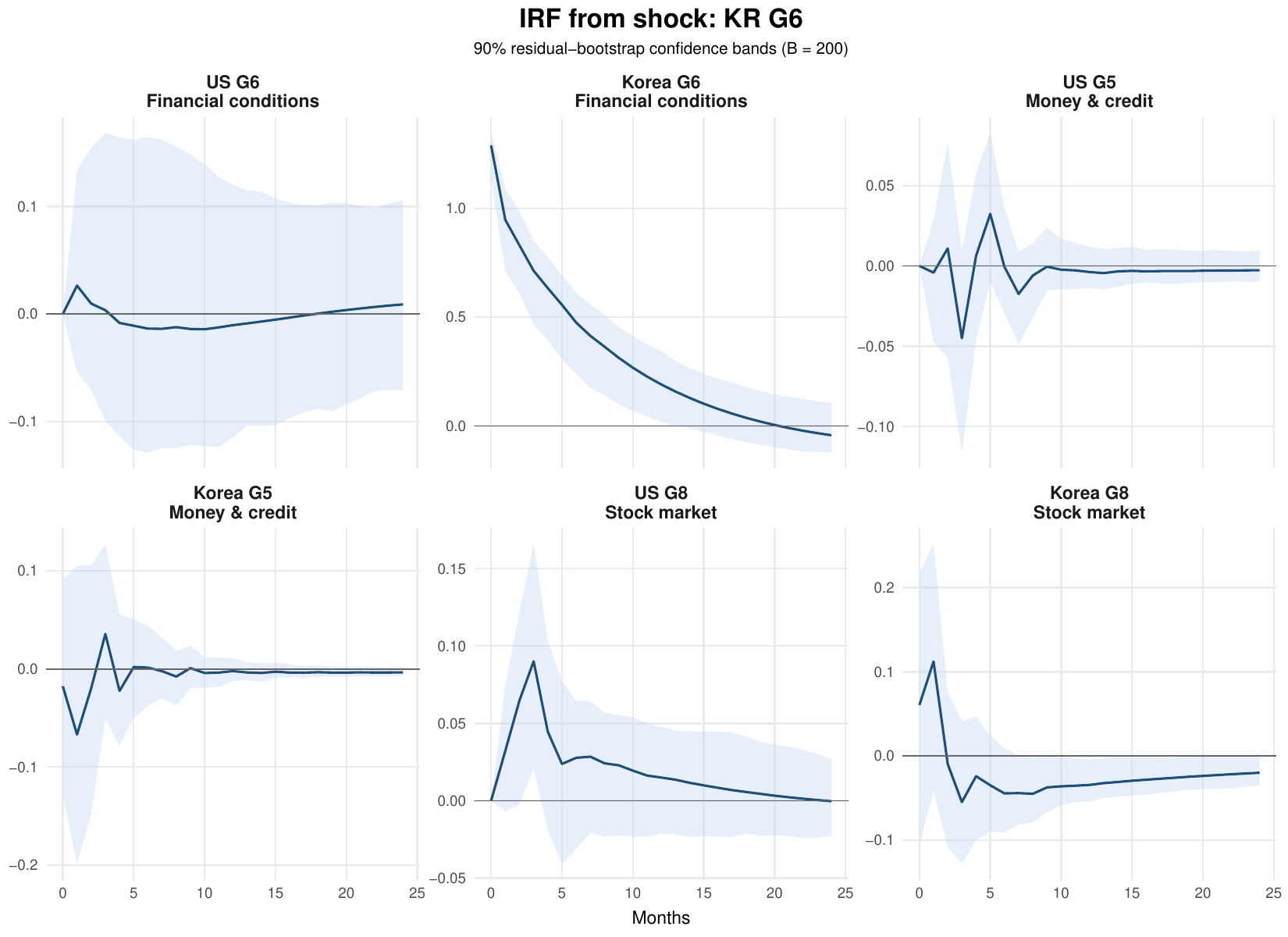}
	\caption{Orthogonalized impulse responses under the U.S.-first financial ordering. The upper panel shows responses to a one-standard-deviation U.S. financial shock ($\mathrm{US\_G6}$), and the lower panel shows responses to the corresponding Korean financial shock ($\mathrm{KR\_G6}$). Shaded regions denote 90\% residual-bootstrap confidence bands based on 200 replications.}
	\label{f:tenar-oirf-financial}
\end{figure}

Figure~\ref{f:tenar-oirf-financial} presents orthogonalized impulse responses under the U.S.-first financial ordering and complements the coefficient matrices in Figure~\ref{f:tenar-dependence-matrix}. The coefficient matrices summarize lag-specific linear dependence, whereas the impulse responses trace how orthogonalized shocks propagate through the full dynamic system. The confidence bands in this tensor application are based on a separate residual bootstrap with 200 replications and are therefore distinct from the Procedure A inference used for the FAVAR in Section~\ref{s:favar}. Under the adopted ordering, the strongest cross-country transmission occurs through the financial-conditions block $G6$. A positive U.S. financial shock produces a sizable and persistent response in the Korean financial factor, whereas the U.S. financial block responds only modestly to a Korean financial shock. The money-and-credit responses in $G5$ are smaller and short-lived and, unlike those in $G6$, do not exhibit a stable cross-country sign pattern. They are therefore best interpreted as secondary adjustment margins rather than as the primary propagation channel. The stock-market block $G8$ displays a similar but weaker asymmetry. Following a positive U.S. financial shock, the U.S. equity factor declines markedly, whereas the Korean equity response is smaller and less persistent. By contrast, a Korean financial shock generates little systematic response in the U.S. equity factor. The grouped tensor model therefore indicates that, under the adopted ordering, directional propagation runs primarily from the U.S. financial block to the Korean financial block, with weaker and less persistent adjustments in money, credit, and equity markets. The combined coefficient and impulse-response evidence thus distinguishes broad but largely nondirectional real-side linkages from sharper U.S.-to-Korea transmission concentrated in financial conditions.

\section{Conclusion} \label{s:conclusion}

KRED consolidates 125 Korean macroeconomic and financial series in a transparent monthly format aligned with FRED-MD. The four common factors constructed from the leading principal-component space summarize interest rates and term spreads, industrial activity, housing permits, and the labor market. Their factor-based diffusion indices show that these components follow distinct cyclical and medium-run paths, allowing researchers to separate developments that would otherwise be compressed into a single aggregate measure.

The monetary-policy application demonstrates the value of the KRED information set. U.S. tightening is followed by a persistent decline in monthly real GDP and an increase in the Korean policy rate. More importantly, the BBE-FAVAR substantially attenuates the positive CPI price-level response generated by the two-lag small VAR benchmark: the reductions in the average and peak responses are economically sizable, and their primary 90\% confidence intervals exclude zero under the baseline Procedure A, dependent wild, and block wild bootstrap designs.

The grouped U.S.--Korea analysis provides a complementary cross-country result. Under the adopted ordering, the strongest and most persistent international propagation is concentrated in financial conditions, with a sizable response in Korea to a U.S. financial shock and a much weaker response in the reverse direction. Real-side responses are smaller and less persistent. These results demonstrate how a consistently structured Korean macroeconomic panel can clarify domestic policy transmission and the financial channels through which global conditions affect a small open economy.

\small
\section*{Data availability}
The KRED data repository is publicly available at \url{https://github.com/crbaek/KRED}. The repository contains the database files, documentation, and update scripts, and it will be updated regularly.

\section*{Acknowledgments}
This research was initiated during the first author's visit to Professor M. C. D\"uker in February 2025 at the Friedrich-Alexander University of Erlangen-Nuremberg. The authors are grateful for her generous hospitality and the intellectual inspiration that helped shape the early development of this project. The authors also thank Professor Vladas Pipiras at the University of North Carolina at Chapel Hill for his valuable comments, which substantially improved the quality and clarity of the paper.

{\footnotesize
		\renewcommand{\baselinestretch}{.8}
	\setlength{\bibsep}{4pt}
	\bibliography{KRED-bib-v12}
}

\clearpage
\appendix
\normalsize
\section*{Appendix}
\renewcommand{\thesubsection}{\Alph{subsection}}
\numberwithin{equation}{subsection} 

\subsection{Denton method} \label{apdx:denton-method}

We follow Denton’s benchmarking framework \citep{denton1971adjust} on a quarterly, seasonally adjusted series $\mathbf{x}=\{x_t:t\in[T]\}$, where for notational convenience $[T]:=\{1,\ldots,T\}$ and $x_t$ is observed only at year-end quarters $\mathcal{T}_Y:=\{t\in[T]:{\rm mod}_4(t)=0\}$. Let $\mathbf{r}=\{r_t:t\in[T]\}$ be the observed seasonally adjusted quarterly growth rates and define the implied cumulative index $\mathbf{i}=\{i_t:t\in[T]\}$ by 
$$ i_t:=\prod_{s=1}^t (1+r_s). $$ 
Writing $x_t=a_t\,i_t$ introduces a smooth ratio process $\mathbf{a}=\{a_t:t\in[T]\}$ that we estimate by penalizing quarter-to-quarter variation subject to matching the observed benchmarks at $\mathcal{T}_Y$:
\[
\min_{\mathbf{a}}\ \sum_{t=2}^T (a_t-a_{t-1})^2
\quad\text{subject to}\quad
a_t\,i_t=x_t,\ \ t\in\mathcal{T}_Y.
\]
In matrix form, it is equivalent to
\[
\min_{\mathbf{a}}\ \tfrac12\,(D\mathbf{a})^\top(D\mathbf{a})
\quad\text{subject to}\quad
(PI)\,\mathbf{a}=\mathbf{x}_Y,
\]
where $D\in\mathbb{R}^{(T-1)\times T}$ is the first-difference operator, $I:=\operatorname{diag}(i_1,\ldots,i_T)$, $P\in\mathbb{R}^{|\mathcal{T}_Y|\times T}$ selects the year-end quarters with entries
\[
P_{jt}=\begin{cases}
1,& t=4j,\\[2pt]
0,& \text{otherwise},
\end{cases}
\]
and $\mathbf{x}_Y\in\mathbb{R}^{|\mathcal{T}_Y|}$ stacks the observed $\{x_t:t\in\mathcal{T}_Y\}$ chronologically. The KKT conditions yield
\begin{align}\label{kkt}
\begin{pmatrix}
D^\top D & (PI)^\top\\
PI & 0
\end{pmatrix}
\begin{pmatrix}
\mathbf{a}\\
\boldsymbol{\lambda}
\end{pmatrix}
=
\begin{pmatrix}
\mathbf{0}\\
\mathbf{x}_Y
\end{pmatrix},
\end{align}
with Lagrange multipliers $\boldsymbol{\lambda}$. Solving \eqref{kkt} gives $\widehat{\mathbf{a}}$, and the imputed quarterly path is 
$$\widehat{x}_t=\widehat{a}_t\,i_t $$
for $t\in[T]$.

\clearpage
\subsection{Additional tensor impulse responses}\label{apdx:tensor-irf}

\begin{figure}[ht!]
    \centering
    \includegraphics[width=.80\textwidth,keepaspectratio]{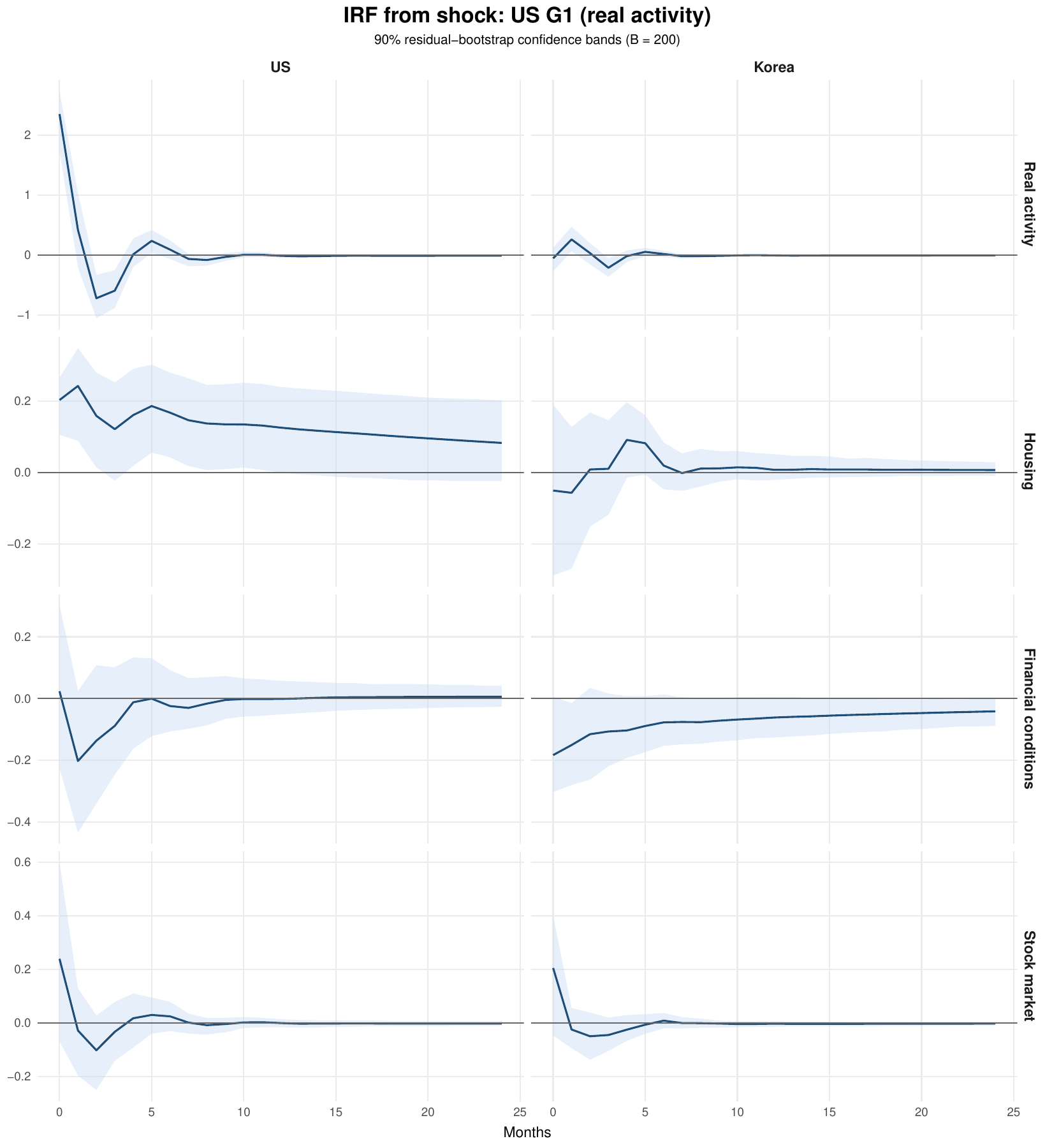}
    \caption{Supplementary generalized impulse responses to a one-standard-deviation U.S. real-activity shock ($\mathrm{US\_G1}$). Shaded regions denote 90\% residual-bootstrap confidence bands based on 200 replications.}
    \label{f:appendix-tensor-usg1}
\end{figure}

Figure~\ref{f:appendix-tensor-usg1} reports a supplementary generalized impulse-response exercise for a one-standard-deviation U.S. real-activity shock ($\mathrm{US\_G1}$). This figure provides a nonfinancial contrast to the orthogonalized financial responses in Section~\ref{s:tenar}. Relative to the financial responses, the cross-country reactions are smaller, less persistent, and more sensitive to the horizon. This contrast reinforces the interpretation that international dependence in the grouped U.S.--Korea data is concentrated in financial conditions and related market blocks rather than uniformly across all macroeconomic sectors.

\clearpage
\subsection{FAVAR residual diagnostics and bootstrap robustness}\label{apdx:favar-robustness}

The Kilian-style correction uses 300 preliminary bootstrap draws, while the main confidence intervals use 3000 replications. The Hall-recentered intervals are based on centered bootstrap impulse-response errors; the Efron percentile intervals use the empirical quantiles of the bootstrap responses directly. Under Procedure A, the factors and the FAVAR are re-estimated in every bootstrap sample.

The idiosyncratic residuals provide a direct reason to examine dependence-robust multipliers. At conventional Ljung--Box levels, 87.25\% of the idiosyncratic residual series reject no serial correlation, and 65.69\% reject no serial dependence in squared residuals. The mean absolute contemporaneous cross-correlation is 0.113 and its 95th percentile is 0.312. The DWB and BWB designs preserve the full cross-sectional residual vector while allowing the time multiplier to be dependent. A common scalar multiplier is applied to the full residual vector at each date or block, while the state-VAR and idiosyncratic multiplier sequences are generated independently.

Table~\ref{tab:bootstrap-stability} summarizes numerical stability. All 3000 main replications for the preferred two-lag specification return numerical output. A small fraction of re-estimated bootstrap systems is dynamically unstable, but the rate remains around 4.4--5.8\% across the baseline and dependence-robust designs. The 13-lag systems are much less stable, with unstable shares ranging from roughly one quarter to more than one half under some resampling variants. This difference is one reason the 13-lag results are treated as a sensitivity check.

\begin{table}[ht!]
\centering
\caption{Bootstrap re-estimation stability.}
\small
\begin{tabular}{lcc}
\toprule
Lag specification & Bootstrap design & Unstable re-estimates\\
\midrule
$p=2$  & Baseline Procedure A & 5.8\%\\
$p=2$  & DWB/BWB variants & approximately 4.4--5.7\%\\
$p=13$ & Baseline Procedure A & 28.6\%\\
$p=13$ & DWB/BWB full-system variants & approximately 25.1--52.5\%\\
$p=13$ & Paired variants & approximately 24.3--52.1\%\\
\bottomrule
\end{tabular}
\label{tab:bootstrap-stability}
\end{table}

Table~\ref{tab:price-puzzle-hall-p13} reports the primary intervals for the 13-lag paired comparison. Every interval contains zero, consistent with the absence of a significant paired VAR--FAVAR contrast under the 13-lag sensitivity specification.

\begin{table}[ht!]
\centering
\caption{Primary 90\% bootstrap intervals for paired CPI price-level attenuation, $p=13$. Interval endpoints are reported in cumulative log units.}
\small
\begin{tabular}{lcc}
\toprule
Bootstrap design & Mean VAR--FAVAR contrast & Peak-response reduction\\
\midrule
Procedure A & $[-0.000970,\ 0.003388]$ & $[-0.001465,\ 0.002694]$\\
DWB, $L=6$  & $[-0.000693,\ 0.003506]$ & $[-0.001189,\ 0.002784]$\\
DWB, $L=12$ & $[-0.000851,\ 0.003376]$ & $[-0.001512,\ 0.002693]$\\
BWB, $L=6$  & $[-0.000868,\ 0.003452]$ & $[-0.001509,\ 0.003070]$\\
BWB, $L=12$ & $[-0.000895,\ 0.003640]$ & $[-0.001290,\ 0.002721]$\\
\bottomrule
\end{tabular}
\label{tab:price-puzzle-hall-p13}
\end{table}

\begin{figure}[ht!]
\centering
\includegraphics[width=.96\textwidth,keepaspectratio]{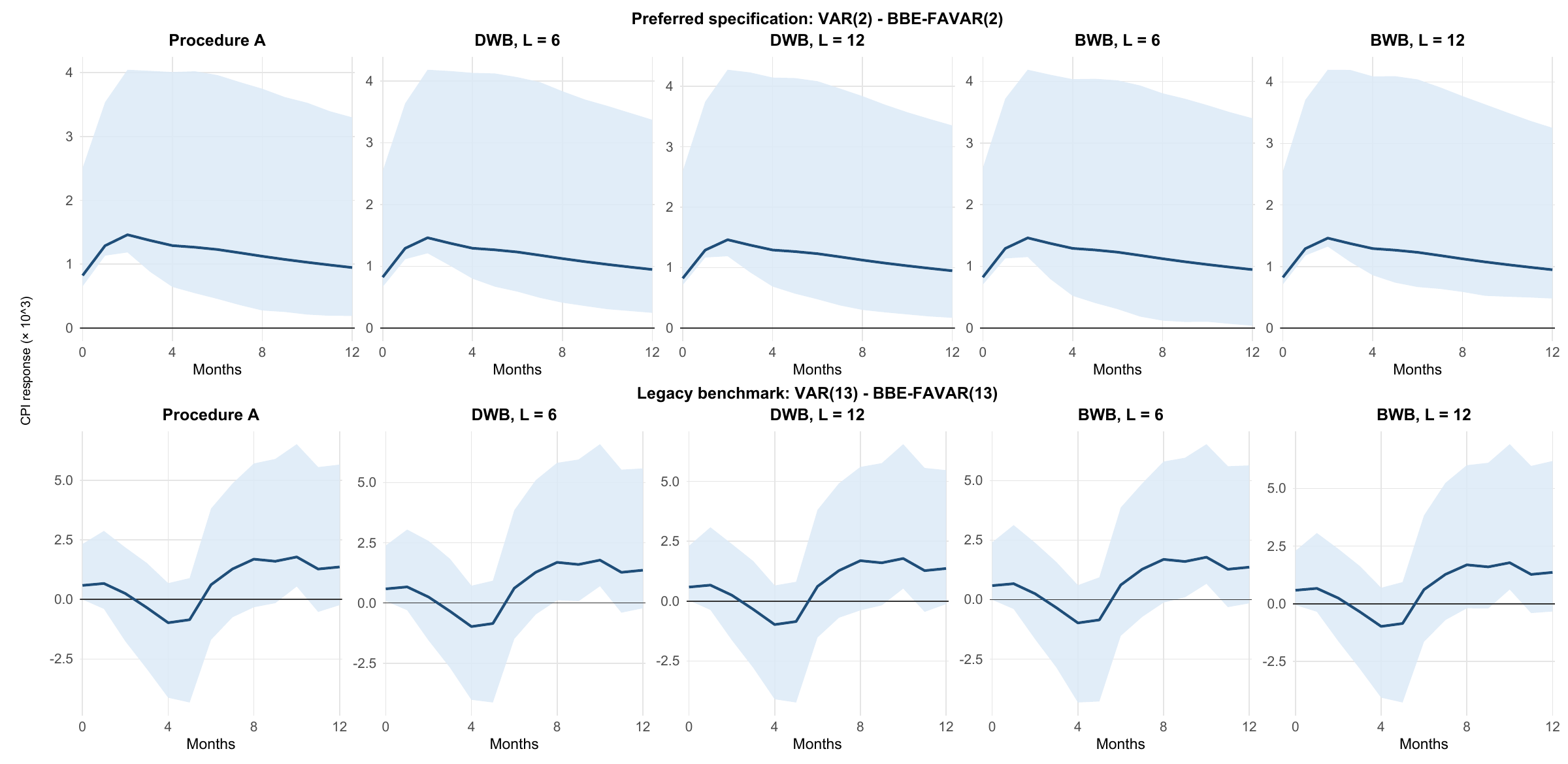}
\caption{Dependence-robust paired CPI price-level contrasts. The figure compares baseline Procedure A, DWB with $L=6,12$, and BWB with $L=6,12$ for the two- and 13-lag specifications.}
\label{f:appendix-price-puzzle-bootstrap}
\end{figure}

Figure~\ref{f:appendix-price-puzzle-bootstrap} complements Tables~\ref{tab:price-puzzle-hall} and~\ref{tab:price-puzzle-hall-p13} by displaying the paired CPI price-level contrasts at each horizon. For the preferred two-lag specification, Procedure A, DWB, and BWB produce closely aligned positive contrast paths, indicating that the attenuation result is not driven by a particular bootstrap design. In the 13-lag specification, the paired response changes sign at short horizons and is surrounded by wider confidence bands, consistent with the statistically insignificant summary contrasts in Table~\ref{tab:price-puzzle-hall-p13}.

\clearpage

\begin{landscape}
\subsection{Variable List and KRED/FRED-MD mapping} \label{appendix:KREDvariables}
We use the transformation-code convention of \citet{mccracken2015fredmd}. The code denotes the following data transformation for a series $x$: (1) no transformation; (2) $\Delta x_t$, (3)
$\Delta^2 x_t$; (4) $\log(x_t)$; (5) $\Delta \log (x_t)$; (6) $\Delta^2 \log(x_t)$; (7) $\Delta(x_t/x_{t-1}-1)$.
\begin{table}[htbp]
	\centering
	\scriptsize
	\caption{Group 1: Output and Income}
	\begin{tabular}{ccp{2cm}p{6cm}p{12cm}} \hline
		id    & tcode & \multicolumn{1}{c}{Name} & \multicolumn{1}{c}{FRED description} & \multicolumn{1}{c}{KRED description} \\ \hline \hline
		
		1  & 5 & \texttt{GDP\_real}        & \texttt{RPI}; Real Personal Income
		& Proxy series; Seasonally adjusted quarterly real GDP constructed from  (ECOS 2.1.1.1: Gross domestic product (annual) \& GDP deflator (annual) \& 2.1.1.2: Growth rates by economic activities --- Gross domestic product (quarterly, SA)) via the Denton method. \\
		
		2  & 5 & \texttt{GNI\_real}    & 
		\texttt{W875RX1}; Real Personal Income ex transfer receipts
		& Proxy series; Seasonally adjusted quarterly real GNI constructed from  (ECOS 2.1.1.1: Gross national income (annual) \& GDP deflator (annual) \& 2.1.1.2: Growth rates of gross national income --- Seasonally adjusted series --- Nominal GNI (quarterly, SA)) via the Denton method. \\
		
		3  & 5 & \texttt{INDPRO}     & IP Index                                                & KOSIS Mining and Manufacturing Production Index (2020=100, SA) \\
		4  & 5 & \texttt{IPFPNSS1}   & IP: Final Products and Nonindustrial Supplies          & KOSIS Index of All Industry Production — Construction (2020=100, SA) \\
		5  & 5 & \texttt{IPFPNSS2}   & IP: Final Products and Nonindustrial Supplies          & KOSIS Index of All Industry Production — Service Industry (2020=100, SA) \\
		
		6  & 5 & \texttt{IPFINAL1}   & IP: Final Products (Market Group)                      & KOSIS Production index, Whole country, Capital goods, (2020=100, SA)  \\
		7  & 5 & \texttt{IPFINAL2}   & IP: Final Products (Market Group)                      & KOSIS Production index, Whole country, Consumers' goods, (2020=100, SA)  \\
		
		8  & 5 & \texttt{IPCONGD}    & IP: Consumer Goods                                     & ECOS 8.3.2; Production Index by Product Group — Consumers' Goods (2020=100, SA) \\
		9  & 5 & \texttt{IPDCONGD}   & IP: Durable Consumer Goods                             & ECOS 8.3.2; Production Index — Durable Consumers' Goods (2020=100, SA) \\
		10  & 5 & \texttt{IPNCONGD}   & IP: Nondurable Consumer Goods                          & ECOS 8.3.2; Production Index — Nondurable Consumers' Goods (2020=100, SA) \\
		11  & 5 & \texttt{IPBUSEQ}    & IP: Business Equipment                                 & ECOS 8.3.3; Machinery production index (2020=100, SA) \\
		12 & 5 & \texttt{IPMAT}      & IP: Materials                                          & ECOS 8.3.2;
        Production Index — Intermediate Goods (2020=100, SA) \\
		13 & 5 & \texttt{IPDMAT}    & IP: Durable Materials                                  & KOSIS Production Index — Durable consumers' goods (2020=100, SA) \\
		14 & 5 & \texttt{IPNMAT}    & IP: Nondurable Materials                               & KOSIS Production Index — Non-durable consumers' goods (2020=100, SA)  \\
		15 & 5 & \texttt{IPMANSICS}  & IP: Manufacturing (SIC)                                & KOSIS Mining and Manufacturing IP — Manufacturing (2020=100, SA) \\
		16 & 5 & \texttt{IPB51222S}  & IP: Residential Utilities                              & KOSIS Mining and Manufacturing IP — Electricity, Gas and Steam supply (2020=100, SA) \\
		17 & 5 & \texttt{IPFUELS}    & IP: Fuels                                              & ECOS 8.3.2; Production Index — Fuel and Electricity (2020=100, SA) \\
		18 & 5 & \texttt{CUMFNS}     & Capacity Utilization: Manufacturing                     & KOSIS Index of manufacturing capacity utilization rate (2020=100, SA) \\
		
		\hline
	\end{tabular}%
	\label{tab:gp1}%
\end{table}%

\end{landscape}

\clearpage
\begin{landscape}
\begin{table}[ht!]
	\centering
	\scriptsize
	\caption{Group 2: Labor Market}
	\begin{tabular}{ccp{2.5cm}p{6.5cm}p{11cm}} \hline
		id    & tcode & \multicolumn{1}{c}{Name} & \multicolumn{1}{c}{FRED description} & \multicolumn{1}{c}{KRED description} \\ \hline \hline
		
		19 & 5 & \texttt{HWI}        & Help Wanted Index
		& Proxy series for vacancy conditions. Monthly newly registered job openings; The Ministry of Employment and Labor (Work-Net: Employment Stability Information Network, EIS: Employment Information Integration and Analysis System) \\
		
		20 & 2 & \texttt{HWIURATIO}  & Ratio of Help Wanted / No. Unemployed
		& Job openings-to-seekers ratio ({\it guinbaesu} in Korean, The Ministry of Employment and Labor; \url{index.go.kr}). \\
		
		21 & 5 & \texttt{CLF16OV}    & Civilian Labor Force & KOSIS Summary of economically active pop. by gender — Labor Force Participation rate (\%) \\
		22 & 5 & \texttt{CE16OV}     & Civilian Employment  & KOSIS Summary of economically active pop. by gender — Employed persons Total (Unit: Thousand Person) \\
		23 & 2 & \texttt{UNRATE}     & Civilian Unemployment Rate & KOSIS Summary of economically active pop. by gender — Unemployment Rate (\%) \\
		24 & 2 & \texttt{UEMPMEAN}   & Average Duration of Unemployment (Weeks) & Unemployment in Months is calculated from (\ref{e:uempmean-formula}) \\
		25 & 5 & \texttt{UEMPLT5}    & Civilians Unemployed — Less Than 5 Weeks & KOSIS Unemployed persons by duration of seeking for work —  12 months and over (Unit: Thousand Person) \\
		26 & 5 & \texttt{UEMP5TO14}  & Civilians Unemployed for 5–14 weeks & KOSIS Unemployed persons by duration of seeking for work — Less than 3 months (Unit: Thousand Person) \\
		27 & 5 & \texttt{UEMP15OV}   & Civilians Unemployed $\ge$ 15 weeks & KOSIS Unemployed persons by duration of seeking for work — 3 months and over (sum of ``3 months and over \& less than 6 months'' and ``6 months and over'') (Unit: Thousand Person) \\
		28 & 5 & \texttt{UEMP15T26}  & Civilians Unemployed for 15–26 weeks & KOSIS Unemployed persons by duration of seeking for work — 3 months and over \& less than 6 months (Unit: Thousand Person) \\
		29 & 5 & \texttt{UEMP27OV}   & Civilians Unemployed $\ge$ 27 weeks & KOSIS Unemployed persons by duration of seeking for work — 6 months and over (Unit: Thousand Person) \\
		30 & 5 & \texttt{PAYEMS}     & All Employees: Total nonfarm & KOSIS Summary of economically active pop. by gender — Employed persons Non-farm household (Unit: Thousand Person) \\
		31 & 5 & \texttt{ICSA}       & Initial Claims & Korea Employment Information Service Employment Administration Statistics  — Labor Market Status — Unemployment Benefit Claims (Monthly) \\
		32 & 5 & \texttt{USGOOD}     & All Employees: Goods-Producing Industries & Employment and Labor Statistics Portal; All Employees: Goods-Producing Industries (BCF) \\
		33 & 5 & \texttt{CES1021000001} & All Employees: Mining and Logging: Mining & Employment and Labor Statistics Portal; All Employees: Construction (B) \\
		34 & 5 & \texttt{USCONS}     & All Employees: Construction & Employment and Labor Statistics Portal; All Employees: Construction (F) \\
		35 & 5 & \texttt{MANEMP}     & All Employees: Manufacturing & Employment and Labor Statistics Portal; All Employees: Manufacturing (C) \\
		36 & 5 & \texttt{DMANEMP}    & All Employees: Durable goods & Employment and Labor Statistics Portal; All Employees: Durable goods (C16, C23, C24, C25, C26, C27, C28, C29, C30, C31, C32, C33) \\
		37 & 5 & \texttt{NDMANEMP}   & All Employees: Nondurable goods & Employment and Labor Statistics Portal; All Employees: Nondurable goods (C10, C11, C12, C13, C14, C15, C17, C18, C19, C20, C21, C22) \\
		38 & 5 & \texttt{SRVPRD}     & All Employees: Service-Providing Industries & Employment and Labor Statistics Portal; All Employees: Service-Providing Industries (G$\sim$S) \\
		39 & 5 & \texttt{USTPU}      & All Employees: Trade, Transportation \& Utilities & Employment and Labor Statistics Portal; All Employees: Trade, Transportation \& Utilities (GHDE) \\
		\hline
	\end{tabular}%
	\label{tab:group2}%
\end{table}%
\end{landscape}

\clearpage
\begin{landscape}

\begin{table}[t!]
	\centering
	\scriptsize
	\begin{tabular}{ccp{2.5cm}p{6cm}p{11cm}} \hline
		id    & tcode & \multicolumn{1}{c}{Name} & \multicolumn{1}{c}{FRED description} & \multicolumn{1}{c}{KRED description} \\ \hline \hline
		
		40 & 5 & \texttt{USWTRADE}   & All Employees: Wholesale Trade & Employment and Labor Statistics Portal; Number of Employed Persons in Wholesale and Merchandise Brokerage, not Wholesale Trade \\
		41 & 5 & \texttt{USTRADE}    & All Employees: Retail Trade & Unlike the United States, the retail trade sector excludes automobile-related industries in Korea (Retail Trade) \\
		42 & 5 & \texttt{USFIRE}     & All Employees: Financial Activities & Employment and Labor Statistics Portal; All Employees: Financial Activities (KL) \\

		43 & 5 & \texttt{USGOV}      & All Employees: Government & Employment and Labor Statistics Portal; All Employees: Government (OPQ) \\
		44 & 2 & \texttt{CES0600000007} & Avg Weekly Hours: Goods-Producing & Employment and Labor Statistics Portal; All Employees: Wages and Working Hours by Industry and Scale; Total hours worked; Goods-Producing (BCF) \\
		45 & 2 & \texttt{AWOTMAN}    & Avg Weekly Overtime Hours: Manufacturing & Avg Weekly Overtime Hours: Manufacturing \& Employment and Labor Statistics Portal; Wages and Working Hours by Industry and Scale; Overtime hours worked of permanent employees: Manufacturing (C) \\
		46 & 2 & \texttt{AWHMAN}     & Avg Weekly Hours: Manufacturing & Avg Weekly Hours: Manufacturing \& Employment and Labor Statistics Portal; Wages and Working Hours by Industry and Scale; Total hours worked: Manufacturing (C) \\
		47 & 6 & \texttt{CES0600000008} & Avg Hourly Earnings: Goods-Producing & Avg Hourly Earnings: Goods-Producing \& Employment and Labor Statistics Portal; Wages and Working Hours by Industry and Scale; Regular wages of permanent employees: Goods-Producing (BCF) \\
		48 & 6 & \texttt{CES2000000008} & Avg Hourly Earnings: Construction & Avg Hourly Earnings: Construction \& Employment and Labor Statistics Portal; Wages and Working Hours by Industry and Scale; Regular wages of permanent employees: Construction (F) \\
		49 & 6 & \texttt{CES3000000008} & Avg Hourly Earnings: Manufacturing & Avg Hourly Earnings: Manufacturing \& Employment and Labor Statistics Portal; Wages and Working Hours by Industry and Scale; Regular wages of permanent employees: Manufacturing (C) \\
		\hline
	\end{tabular}%
	\label{tab:group2-continued}%
\end{table}%

\begin{table}[t!]
	\centering
	\small
	\caption{Group 3: Housing}
	\begin{tabular}{ccp{2.5cm}p{4.5cm}p{12cm}} \hline
		id    & tcode & \multicolumn{1}{c}{Name} & \multicolumn{1}{c}{FRED description} & \multicolumn{1}{c}{KRED description} \\ \hline \hline
		50 & 4 & \texttt{HOUST}    & Housing Starts: Total New Privately Owned & KOSIS; Housing Construction Statistical; Commencement of Housing Construction by Housing Type (Households Monthly Total) \\
		51 & 4 & \texttt{HOUSTNE}  & Housing Starts, Northeast                 & KOSIS; Seoul \\
		52 & 4 & \texttt{HOUSTMW}  & Housing Starts, Midwest                   & KOSIS; Incheon/Gyeonggi \\
		53 & 4 & \texttt{HOUSTS}   & Housing Starts, South                    & KOSIS; 5 local major cities (Busan/Daegu/Ulsan/Gwangju/Daejeon) \\
		54 & 4 & \texttt{HOUSTW}   & Housing Starts, West                      & KOSIS; Others \\
		55 & 4 & \texttt{PERMIT}   & New Private Housing Permits (SAAR)        & KOSIS; Housing Construction Statistical; Statistics of Housing Construction Permits by Category (Monthly total) \\
		56 & 4 & \texttt{PERMITNE} & SAAR–Northeast                            & KOSIS; Seoul \\
		57 & 4 & \texttt{PERMITMW} & SAAR–Midwest                              & KOSIS; Incheon/Gyeonggi \\
		58 & 4 & \texttt{PERMITS}  & SAAR–South                                & KOSIS; 5 local major cities (Busan/Daegu/Ulsan/Gwangju/Daejeon) \\
		59 & 5 & \texttt{PERMITW}  & SAAR–West                                 & KOSIS; Others \\
		\hline
	\end{tabular}%
	\label{tab:gp3}%
\end{table}%

\end{landscape}
\clearpage

\begin{landscape}
\begin{table}[t!]
	\scriptsize
	\centering
	\caption{Group 4: Consumption, orders and inventories}
	\begin{tabular}{ccp{3cm}p{4.5cm}p{12cm}} \hline
		id    & tcode & \multicolumn{1}{c}{Name} & \multicolumn{1}{c}{FRED description} & \multicolumn{1}{c}{KRED description} \\ \hline \hline
		
		60 & 5 & \texttt{DPCERA3M086SBEA} & Real personal consumption expenditures
		& ECOS 2.1.7.1.2; Final Consumption Expenditure of Resident Households by Purpose (seasonally adjusted, chained 2020 year prices, quarterly) \\
		
		-- & -- & \texttt{CMRMTSPx} & Real Manu. And Trade Industries Sales
		& Not included in KRED (available in ECOS only at annual frequency; ECOS; 5.1.3.1. Relationship Ratios of Income and Expenses(11th revised KSIC) -- All industry). \\
		
		61 & 5 & \texttt{RETAILx} & Retail and Food Services Sales
		& KOSIS Retail and Food Services Sales (2020=100.0) (Constant Index) \\
		62 & 2 & \texttt{ACOGNO} & New Orders for Consumer Goods
		& Proxy based on the KOSIS Manufacturing Domestic Supply Index (consumer goods, domestic supply), not a direct new-orders series. \\
		63 & 5 & \texttt{AMDMNOx} & New Orders for Durable Goods
		& KOSIS Value of machinery order  received by kind of consumer/product(current) — Total value ordered - Vessels (Unit : Million won)\\
		64 & 5 & \texttt{ANDENOx} & New Orders for Nondefense Capital Goods
		& KOSIS Value of machinery order  received by kind of consumer/product(current) — Total value ordered (Unit : Million won) - Excluding vessels (Unit : Million won) \\
		65 & 5 & \texttt{AMDMUOx} & Unfilled Orders for Durable Goods
		& KOSIS Value of machinery order  received by kind of consumer/product(current) — Outstanding orders (Unit : Million won)\\
		-- & -- & \texttt{BUSINVx} & Total Business Inventories
		& Not included in KRED (no monthly analogue available). \\
		66 & 2 & \texttt{ISRATIOx} & Total Business: Inventories to Sales Ratio
		& ECOS 8.3.5; Index of Inventory Turnover Ratio -- Manufacturing (Unit : 2020 = 100) \\
		67 & 2 & \texttt{UMCSENTx} & Consumer Sentiment Index
		& ECOS 6.2.1; Consumer Tendency Survey -- Composite Consumer Sentiment Index (BOK, National) (Monthly) \\
		\hline
	\end{tabular}%
	\label{tab:gp4}%
\end{table}%

\begin{table}[t!]
	\scriptsize
	\centering
	\caption{Group 5: Money and credit}
	\begin{tabular}{ccp{2.6cm}p{4.5cm}p{12cm}} \hline
		id    & tcode & \multicolumn{1}{c}{Name} & \multicolumn{1}{c}{FRED description} & \multicolumn{1}{c}{KRED description} \\ \hline \hline
		68 & 6 & \texttt{M1SL}        & M1 Money Stock & ECOS 1.1.2.1.2; M1 By Type (Average, Unit : Bil.Won) \\
		69 & 6 & \texttt{M2SL}        & M2 Money Stock & ECOS 1.1.3.1.2; M2 By Type (Average, Unit : Bil.Won) \\
		70 & 5 & \texttt{M2REAL}      & Real M2 Money Stock & ECOS; \texttt{M2SL} divided by \texttt{CPIAUCSL}, where \texttt{CPIAUCSL} denotes the CPI (2020=100) \\
		71 & 6 & \texttt{BOGMBASE}    & Monetary Base & ECOS 1.1.1.1.4; Components of Monetary Base (End of, Unit : Bil.Won) \\
		72 & 6 & \texttt{TOTRESNS}    & Total Reserves of Depository Institutions & ECOS 1.4.3.1; Reserves of Commercial and Specialized Banks (New version, average of, Unit : Mil.Won) \\
		73 & 7 & \texttt{NONBORRES}   & Nonborrowed Reserves & ECOS 1.4.1; Reserves Deposits of BFCs - (Loans \& Discount + Loans to Govt) (Unit : Bil. Won) \\
		74 & 6 & \texttt{BUSLOANS}    & Commercial and Industrial Loans & ECOS 1.2.5.1.1; All Industries except Construction and Real Estate Activities in Service (Unit : Bil. Won) \\
		75 & 6 & \texttt{REALLN}      & Real Estate Loans at All Commercial Banks & ECOS 1.2.5.1.1; Construction + 1.2.5.1.1; Real Estate Activities in Service + 1.2.4.1.2; Household mortgage loans by Depository Corporations (Unit : Bil. Won)\\
		76 & 6 & \texttt{NONREVSL}    & Total Nonrevolving Credit & ECOS 1.2.4.1.2; By Purpose - Others by Depository Corporations \\
		77 & 2 & \texttt{CONSPI}      & Nonrevolving consumer credit to Personal Income & ECOS; \texttt{NONREVSL}/\texttt{GDP\_real} \\
		78 & 6 & \texttt{DTCOLNVHFNM} & Consumer Motor Vehicle Loans Outstanding & MOLIT; Total Registered Motor Vehicles (Private Use) \\
		79 & 6 & \texttt{DTCTHFNM}    & Total Consumer Loans and Leases Outstanding & ECOS 1.2.4.2.1; Deposits, Loans \& Discounts By Section (Unit : Bil.Won) \\
		80 & 6 & \texttt{INVEST}      & Securities in Bank Credit at All Commercial Banks & ECOS 1.1.6.1; Depository Corporations Survey (end of period); sum of \texttt{BBABAA1011}, \texttt{BBABAA201}, \texttt{BBABAA304}, \texttt{BBABAA402}, \texttt{BBAAAA602}, \texttt{BBABAB102}, \texttt{BBABAA403}, \texttt{BBAAAA604}, and \texttt{BBABAB108}
 \\
		\hline
	\end{tabular}%
	\label{tab:gp5}%
\end{table}%
\end{landscape}

\clearpage
\begin{landscape}
\begin{table}[t!]
	\centering
	\scriptsize
	\caption{Group 6: Interest rate and Exchange rates}
	\begin{tabular}{ccp{2.5cm}p{5cm}p{12cm}} \hline
		id    & tcode & \multicolumn{1}{c}{Name} & \multicolumn{1}{c}{FRED description} & \multicolumn{1}{c}{KRED description} \\ \hline \hline
		81 & 2 & \texttt{KR.MIR} & FEDFUNDS/Effective Federal Funds Rate
		& ECOS 1.3.2.2; Market Interest Rates (Monthly, Quarterly, Annually); Uncollateralized Call Rates (Overnight) (Percent Per Annum) \\
		82 & 2 & \texttt{CP3Mx} & 3-Month AA Financial Commercial Paper Rate
		& ECOS 1.3.2.2; Yields on CP (91-day) (Percent Per Annum) \\
		83 & 2 & \texttt{TB3MS} & 3-Month Treasury Bill Secondary Market Rate, Discount Basis
		& ECOS 1.3.2.2; Market Interest Rates (Monthly, Quarterly, Annually); Monetary stabilization bonds (91-day) \\
		84 & 2 & \texttt{TB6MS} & 6-Month Treasury Bill
		& ECOS 1.3.2.2; Market Interest Rates (Monthly, Quarterly, Annually); Yields of Monetary Stab. Bonds (1-year) \\
		85 & 2 & \texttt{GS1} & 1-Year Treasury Rate
		& ECOS 1.3.2.2; Yields of Treasury Bonds (1-year) (Percent Per Annum) \\
		86 & 2 & \texttt{GS5} & 5-Year Treasury Rate
		& ECOS 1.3.2.2; Yields of Treasury Bonds (3-year) (Percent Per Annum) \\
		87 & 2 & \texttt{GS10} & 10-Year Treasury Rate
		& ECOS 1.3.2.2; Yields of Treasury Bonds (10-year) (Percent Per Annum) \\
		88 & 2 & \texttt{AAA} & Moody’s Seasoned Aaa Corporate Bond Yield
		& ECOS 1.3.2.2; Yields of Corporate Bonds: O.T.C (3-year, AA-) (Percent Per Annum) \\
		89 & 2 & \texttt{BAA} & Moody’s Seasoned Baa Corporate Bond Yield
		& ECOS 1.3.2.2; Yields of Corporate Bonds: O.T.C (3-year, BBB-) (Percent Per Annum) \\
		90 & 1 & \texttt{COMPAPFFx} & 3-Month Commercial Paper Minus FEDFUNDS
		& ECOS 1.3.2.2; Yields on CP (91-day) $-$ Uncollateralized Call Rates (Overnight) \\
		91 & 1 & \texttt{TB3SMFFM} & 3-Month Treasury C Minus FEDFUNDS
		& ECOS 1.3.2.2; MSB (91-day) minus Uncollateralized Call Rates (Overnight) \\
		92 & 1 & \texttt{TB6SMFFM} & 6-Month Treasury C Minus FEDFUNDS
		& ECOS 1.3.2.2; MSB (1-year) Minus Uncollateralized Call Rates (Overnight) \\
		93 & 1 & \texttt{T1YFFM} & 1-Year Treasury C Minus FEDFUNDS
		& ECOS 1.3.2.2; Yields of Treasury Bonds (1-year) Minus Uncollateralized Call Rates (Overnight) \\
		94 & 1 & \texttt{T3YFFM} & 5-Year Treasury C Minus FEDFUNDS
		& ECOS 1.3.2.2; Yields of Treasury Bonds (3-year) Minus Uncollateralized Call Rates (Overnight) \\
		95 & 1 & \texttt{T10YFFM} & 10-Year Treasury C Minus FEDFUNDS
		& ECOS 1.3.2.2; Yields of Treasury Bonds (10-year) Minus Uncollateralized Call Rates (Overnight) \\
		96 & 1 & \texttt{AAAFFM} & Moody’s Aaa Corporate Bond Minus FEDFUNDS
		& ECOS 1.3.2.2; Yields of Corporate Bonds: O.T.C (3-year, AA-) Minus Uncollateralized Call Rates (Overnight) \\
		97 & 1 & \texttt{BAAFFM} & Moody’s Baa Corporate Bond Minus FEDFUNDS
		& ECOS 1.3.2.2; Yields of Corporate Bonds: O.T.C (3-year, BBB-) Minus Uncollateralized Call Rates (Overnight) \\
		98 & 5 & \texttt{TWEXAFEGSMTH} & Trade Weighted U.S. Dollar Index
		& BIS: Nominal effective exchange rate, Korea / Broad basket (2020=100). \url{https://data.bis.org/topics/EER/BIS,WS_EER,1.0/M.N.B.KR} \\
		99 & 5 & \texttt{EXKRUSx} & Switzerland / U.S. Foreign Exchange Rate
		& ECOS 3.1.2.1; Arbitraged Rates of Major Currencies Against Won, Longer Frequency $-$ Won per United States Dollar (Basic Exchange Rate) (Closing Rate, unit: won) \\
		100 & 5 & \texttt{EXKRJPx} & Japan / U.S. Foreign Exchange Rate
		& ECOS 3.1.2.1; Won per Japanese Yen (100Yen) (Closing Rate, unit: won) \\
		101 & 5 & \texttt{EXKREUx} & U.S. / U.K. Foreign Exchange Rate
		& ECOS 3.1.2.1; Won per Euro (Closing Rate, unit: won) \\
		102 & 5 & \texttt{EXKRCNx} & Canada / U.S. Foreign Exchange Rate
		& ECOS 3.1.2.1; Won per Yuan (Closing Rate, unit: won) \\
		\hline
	\end{tabular}%
	\label{tab:gp6}%
\end{table}%
\end{landscape}

\clearpage
\begin{landscape}
\begin{table}[t!]
	\centering
	\scriptsize
	\caption{Group 7: Prices}
	\begin{tabular}{ccp{3cm}p{6cm}p{11cm}} \hline
		id    & tcode & \multicolumn{1}{c}{Name} & \multicolumn{1}{c}{FRED description} & \multicolumn{1}{c}{KRED description} \\ \hline \hline
		103 & 6 & \texttt{WPSFD49207} & PPI : Finished Goods & ECOS 4.1.2; Domestic Supply Price Indices (2020=100); Final goods and services \\
		104 & 6 & \texttt{WPSFD49502} & PPI : Finished Consumer Goods & ECOS 4.1.2; Domestic Supply Price Indices (2020=100); Consumer goods \\
		105 & 6 & \texttt{WPSID61} & PPI : Intermediate Materials & ECOS 4.1.2; Domestic Supply Price Indices (2020=100); Intermediate goods and services \\
		106 & 6 & \texttt{WPSID62} & PPI : Crude Materials & ECOS 4.1.2; Domestic Supply Price Indices (2020=100); Raw materials \\
		107 & 6 & \texttt{OILPRICEx} & Crude Oil, spliced WTI and Cushing & ECOS 9.1.6.3; World Commodity Prices; Crude oil (Dubai Fateh) (unit : \$ /bbl) \\
		108 & 6 & \texttt{PPICMM} & PPI : Metals and metal products: & ECOS 4.1.1.1; Producer Price Indices (Basic Groups) $-$ Average of Basic metal products \& Fabricated metal products (2020=100） \\
		109 & 6 & \texttt{CPIAUCSL} & CPI : All Items & ECOS 4.2.2; Consumer Price indices (All Cities, Special Groups); All items (2020=100, Wgt : 1000) \\
		110 & 6 & \texttt{CPIAPPSL} & CPI : Apparel & ECOS 4.2.1; Consumer Price indices; Clothing and footwear (2020=100, Wgt : 49.6) \\
		111 & 6 & \texttt{CPITRNSL} & CPI : Transportation & ECOS 4.2.1; Consumer Price indices; Transport (2020=100, Wgt : 110.6) \\
		112 & 6 & \texttt{CPIMEDSL} & CPI : Medical Care & ECOS 4.2.1; Consumer Price indices; Health (2020=100, Wgt : 84) \\
		113 & 6 & \texttt{CUSR0000SAC} & CPI : Commodities & ECOS 4.2.2; Consumer Price indices (All Cities, Special Groups); Commodities (2020=100, Wgt : 447.6) \\
		114 & 6 & \texttt{CUSR0000SAD} & CPI : Durables & ECOS 4.2.2; Consumer Price indices (All Cities, Special Groups); Durable goods (2020=100, Wgt : 73.0) \\
		115 & 6 & \texttt{CUSR0000SAS} & CPI : Services & ECOS 4.2.2; Consumer Price indices (All Cities, Special Groups); Services (2020=100, Wgt : 552.4) \\
		116 & 6 & \texttt{CPIULFSL} & CPI : All Items Less Food & ECOS 4.2.2; Consumer Price indices (All Cities, Special Groups); Excluding Food \& Energy (2020=100, Wgt : 782.2) \\
		117 & 6 & \texttt{CUSR0000SA0L2} & CPI : All Items Less Shelter &  ECOS 4.2.2; Consumer Price indices (All Cities, Special Groups); Rental for housing (2020=100, Wgt : 99.1) (Note that instead of using `CPI: All Items Less Shelter', we include `Rental for housing', which corresponds to the “Shelter” component.) \\
		-- & -- & \texttt{CUSR0000SA0L5} & CPI : All items less medical care
		& Not included in KRED (CPIAUCSL and CPIMEDSL are available only as index series and the expenditure weights needed to remove medical care are not available). \\
		118 & 6 & \texttt{PCEPI} & Personal Cons. Expend. : Chain Index & ECOS 2.1.7.2.1 \& 2.1.7.2.2; Final Consumption Expenditure of resident households, expressed as the ratio of current prices to 2020 chained prices \\
		119 & 1 & \texttt{DDURRG3M086SBEA} & Personal Cons. Exp : Durable goods & ECOS 2.1.7.2.2; Final Consumption Expenditure of households on the territory (Durable goods, Unit : annualized growth rate) \\
		120 & 1 & \texttt{DNDGRG3M086SBEA} & Personal Cons. Exp : Nondurable goods & ECOS 2.1.7.2.2; Final Consumption Expenditure of households on the territory (Non-durable goods, Unit : annualized growth rate) \\
		121 & 1 & \texttt{DSERRG3M086SBEA} & Personal Cons. Exp : Services & ECOS 2.1.7.2.2; Final Consumption Expenditure of households on the territory (Services, Unit : annualized growth rate) \\
		\hline
	\end{tabular}%
	\label{tab:gp7}%
\end{table}%

\begin{table}[t!]
	\centering
	\scriptsize
	\caption{Group 8: Stock market}
	\begin{tabular}{ccp{2.5cm}p{6cm}p{11cm}} \hline
		id    & tcode & \multicolumn{1}{c}{Name} & \multicolumn{1}{c}{FRED description} & \multicolumn{1}{c}{KRED description} \\ \hline \hline
		122 & 5 & \texttt{KOSPI}        & S\&P's Common Stock Price Index: Composite & ECOS 1.5.1.2; KOSPI\_Index (Avg.) \\
		123 & 2 & \texttt{KOSPIdivyield} & S\&P's Common Stock Price Index: Dividend Yield & ECOS 1.5.1.2; KOSPI\_Dividend yield ratio \\
		124 & 5 & \texttt{KOSPIPE}  & S\&P's Common Stock Price Index: Price-Earnings Ratio & ECOS 1.5.1.2; KOSPI\_Price Earnings Ratio \\
		125 & 1 & \texttt{VKOSPI}        & VIX & KRX Market data system \url{https://data.krx.co.kr}; V-KOSPI 200|Options \\
		\hline
	\end{tabular}%
	\label{tab:gp8}%
\end{table}%
\end{landscape}

\end{document}